\documentclass[usenatbib]{mn2e}
\pdfoutput=1
\usepackage{graphicx}
\usepackage{verbatim}
\usepackage{microtype}
\usepackage{txfonts}
\usepackage{times}
\usepackage{color}

\graphicspath{{./}}
\begin{document}

\label{firstpage}

\title[Spectropolarimetry with the GMRT at 610~MHz: A case study of two
SCG fields]{Spectropolarimetry with the Giant Metrewave Radio Telescope at
610~MHz: a case study of two Southern Compact Group fields}

\author[J.~S.\ Farnes, et al.]{J.~S.\ Farnes$^{1,2}$\thanks{email:
  \texttt{jamie.farnes@sydney.edu.au}},
  D.~A.\ Green$^{1}$ and N.~G.\ Kantharia$^{3}$\\
  $^{1}$Cavendish Laboratory, 19 J.J.\ Thomson Avenue, Cambridge, CB3 0HE\\
  $^{2}$Sydney Institute for Astronomy, School of Physics, University of
        Sydney, NSW 2006, Australia\\
  $^{3}$National Centre for Radio Astrophysics (TIFR), Pune University Campus,
        Pune 411 007, India}

\date{Accepted ---}

\pagerange{\pageref{firstpage}--\pageref{lastpage}}

\pubyear{2013}

\maketitle

\begin{abstract}
%
We present 610~MHz spectropolarimetric images from the Giant Metrewave Radio Telescope (GMRT). We discuss the properties of the GMRT's full-polarisation mode in detail and provide a technical characterisation of the instrument's polarisation performance. We show that the telescope can be used for wide-field spectropolarimetry. The application of Rotation Measure Synthesis results in a sensitivity at the level of tens of $\muup$Jy in Faraday space. As a demonstration of the instrument's capabilities, we present observations of two Southern Compact Groups -- the Grus Quartet and USCG~S063. The observations are compared with other radio and infra-red data to constrain the group members' spectral index, polarisation fraction, and Faraday depth. Radio continuum emission is detected from all four galaxies in the Grus Quartet and one galaxy in USCG~S063. Linear polarisation is detected from a circumnuclear starburst ring in NGC~7552, the active nucleus of NGC~7582, and three extended radio galaxies in the background of the Grus Quartet field. These background sources allow for the classification of an FR-I and an X-shaped radio galaxy. No evidence is found for interaction with the intragroup medium in either galaxy group.
\end{abstract}


\section{Introduction}

Polarised radio emission is fundamentally related to the presence of
magnetic fields, and observations of polarisation are the best way of
directly studying quasi-regular fields
\citep[e.g.][]{2013arXiv1302.0889B}. A key issue for polarimetry of
astrophysical sources is removal of effects that alter the properties of
the radiation as it propagates across the Universe and through the
various components of a radio telescope \citep[e.g.][]{HamakerREF}. The
`full-polarisation' mode of the Giant Metrewave Radio Telescope (GMRT)
-- which provides observations of all four cross-correlation products
i.e.\ $RR$, $LL$, $RL$, $LR$ -- has recently become available at 610~MHz. 

The properties of faint polarised sources at frequencies $\le1.0$~GHz are relatively poorly constrained by observations, despite this being the region where Faraday depolarisation effects are expected to dominate \citep[e.g.][]{1966MNRAS.133...67B}. Measurements of polarised emission require a combination of high sensitivity to probe the weak polarised signals at these frequencies, together with consideration of the numerous systematics that can affect wide-field and low-frequency polarised observations.

There are a number of facilities designed for spectropolarimetry at other frequencies: ASKAP will observe at frequencies between 700~MHz to 1.8~GHz \citep[e.g.][]{2008ExA....22..151J}, ATCA at $\ge1.1$~GHz \citep[e.g.][]{2011MNRAS.416..832W}, GALFACTS between 1.2 to 1.5~GHz \citep[e.g.][]{2010ASPC..438..402T}, the JVLA at $>$1.2~GHz \citep[e.g.][]{2009IEEEP..97.1448P}, and LOFAR at $<$230~MHz \citep[e.g.][]{2012arXiv1203.2467A}. The GMRT itself is currently undergoing an upgrade that will further improve upon its science capabilities \citep{GuptaREF}. When the upgrade is complete, the intended nearly-seamless frequency coverage from 150~MHz to 1.5~GHz, with instantaneous bandwidths of 400~MHz, will be centred within a frequency range with limited complementary observational data. Wide-field polarimetric surveys with the GMRT therefore have the potential to fill a crucial gap in our understanding of cosmic magnetic fields \citep[e.g.][]{2012A&A...543A.113B}.

As a demonstration of the instrument's capabilities, we here present wide-field spectropolarimetric observations of two Southern
Compact Groups (SCGs) made at 610~MHz with the GMRT. Galaxy groups are
some of the smallest and most dense associations of galaxies. These
groups are gravitationally bound structures in various stages of
dynamical evolution, from young and spiral-dominated to almost
completely merged. They are amongst the best natural laboratories for
investigating the evolution of the intragroup medium (IGM). It is known that mergers and less
powerful interactions can be a significant contributor to heating of the
IGM via both supernova explosions and the triggering of active galactic
nuclei (AGN) \citep{2007A&A...473..399P}. These data show that
it is possible to make wide-field spectropolarimetric observations with
the GMRT at 610~MHz. Note that \citet{joshipaper} present 610~MHz
polarisation results from the GMRT using the bandwidth-averaged Stokes
$Q$/$U$ observations. Also, \citet{FarnesRMAA} present other 610~MHz
GMRT spectropolarimetric observations of the nearby galaxy M51, which
is depolarised below the sensitivity limit.

Sections~\ref{S:RMS} and \ref{S:GMRT} present, respectively, background
information related to the technique of Rotation Measure synthesis, and details of the GMRT polarisation observations, data reduction
procedures, and a technical characterisation of the instrument's polarisation performance. The results for the case study observations of the two SCGs
are presented and discussed in Section~\ref{S:groupgalaxies}, and for
other sources in the fields in Section~\ref{S:nongroup}, with
conclusions presented in Section~\ref{S:discussion}. Throughout this
paper a $\Lambda$CDM cosmology has been assumed with $\Omega_{M}=0.27$,
$\Omega_{\Lambda}=0.73$, and $H_{0}=71$~km~s$^{-1}$~Mpc$^{-1}$. The
spectral index, $\alpha$, is defined such that $S\propto\nu^{-\alpha}$.
All stated Faraday depths are the observed values -- a rest-frame
correction of $(1+z)^2$ has not been applied. Unless otherwise
specified, all coordinates are equatorial J2000.

\section{Rotation Measure Synthesis}\label{S:RMS}

A key issue for polarimetry of astrophysical sources is removal of effects that
alter the properties of the radiation as it propagates across the
Universe and through the various components of a radio telescope
\citep[e.g.][]{HamakerREF}. One such effect, Faraday rotation, occurs as
linearly polarised radiation travelling through a magnetised plasma
undergoes birefringence. The linear polarisation can be considered as
two counter-rotating circularly polarised components which experience
different refractive indices. Upon exiting the plasma, Faraday rotation
will have caused the electric vector of the incoming linearly polarised
wave to rotate.

Consider a simple model with just one source along the line of sight
with no internal Faraday rotation, and only a single slab of plasma
existing between us and the source. In this case, the electric vector
polarisation angle (EVPA) will be rotated by an amount proportional to
the squared wavelength of the radiation as described by
\begin{equation}
  \label{EVPArotation}
  \Phi_{\rm EVPA} = \Phi_{0} + \textrm{RM}\,\lambda^2 \,,
\end{equation}
where $\Phi_{\rm EVPA}$ is the observed EVPA, $\Phi_{0}$ is the
intrinsic EVPA at the source, and $\lambda$ is the wavelength of the
radiation. The factor of proportionality is known as the rotation
measure (RM).

The measurement of RM allows magnetic fields oriented along the line of
sight to be probed. The interpretation of RM measurements is complicated
as Faraday rotation occurs both within a source and in the intervening
magnetised plasmas between us and the source. For example, Faraday
rotation from the interstellar medium of our own Galaxy (the `Galactic foreground')
corrupts the measurement of weak magnetic fields in extragalactic
sources. In addition, Faraday rotation in the Earth's ionosphere further
modifies polarised signals and inhibits polarisation calibration.

The relationship between EVPA and $\lambda^2$ can be far more
complicated than the simplified linear-dependence that equation
\ref{EVPArotation} suggests. Many astronomical sources contain both
non-thermal and thermal electron populations -- leading to internal
Faraday rotation. Furthermore, multiple Faraday rotating regions may
exist along the line of sight between us and a source. This can lead to
a non-linear dependence between $\Phi_{\rm EVPA}$ and $\lambda^2$, with
the RM effectively becoming a function of wavelength. The separation of
the differing wavelength-dependent contributions to the RM (which exist
at different Faraday depths) is essential to understand cosmic
magnetism. The capabilities of modern correlators allow for a single
observation to output a large number of frequency channels. In
combination with the technique of RM Synthesis
\citep{2005A&A...441.1217B}, it is possible to separate these differing
contributions to the RM, eliminate $n\pi$ ambiguities, and maximise the
sensitivity to polarised emission.

The RM is defined as
\begin{equation}
  \textrm{RM} =
  \frac{\textrm{d}\Phi_{\rm EVPA}(\lambda^2)}{\textrm{d}\lambda^2} \,,
\end{equation}
and now equation \ref{EVPArotation} can be redefined with a factor of
proportionality equal to the Faraday depth, $\phi$, so that
\begin{equation}
  \label{RM2}
  \phi(s) = \frac{e^3}{2 \pi m_{\rm e}^2 c^4}
      \int_{0}^d n_{\rm e} B_{\parallel} \, ds \,,
\end{equation}
where $n_{\rm e}$ is generally the electron number density of the plasma
and $B_{\parallel}$ is the strength of the component of the magnetic
field parallel to the line of sight. The constants $e$, $m_{\rm e}$, and
$c$ are the electronic charge, the mass of the electron, and the speed
of electromagnetic radiation in a vacuum respectively. The integral from
0 to $d$ represents the distance along the line of sight between the
observer and the source.

Now the RM is simply the slope of a $\Phi_{\rm EVPA}$ versus $\lambda^2$
plot, and $\phi$ is defined so that it can vary along the line of sight
and such that a positive value implies a magnetic field pointing towards
the observer. Sources of radiation may now exist at different Faraday
depths along a line of sight. Note that the Faraday depth does not have
a simple relation to physical depth.

To obtain the Faraday depth from our observations, we perform RM Synthesis and obtain the peak in the deconvolved `Faraday dispersion function' which describes the intrinsic polarisation at each Faraday depth \citep{1966MNRAS.133...67B}. In essence,
the mathematical derivation of the RM Synthesis technique allows for
polarisation vectors that are close together in $\lambda^2$-space to be
Fourier inverted into `Faraday space' i.e.\ $\phi$-space. The
polarisation vectors are summed coherently at a number of trial Faraday
depths, with the sensitivity being maximised to polarised emission at
that Faraday depth. For a full description of the technique, see \citet{2005A&A...441.1217B}.

\begin{table*}
\centering
\begin{minipage}{155mm}
\caption{Details of the observations.}\label{tab:obs}
\begin{tabular}{@{}ccccccccc@{}}\hline
 Date & Target & Observation & Time on source & Frequency & Bandwidth &
   Number of & Best resolution\footnote{Using robust weighting.}\\
      &        & code        & /min & /MHz & /MHz & channels & /arcsec \\\hline
 2010 Jan 08 & SCG~0141$-$3429 & 17\_060\_1 & 220 & 610 & 16 & 256 & $9\farcs7 \times 5\farcs6$\\
 2010 Jan 09 & SCG~2315$-$4241 & 17\_060\_2 & 180 & 610 & 16 & 256 & $9\farcs5 \times 4\farcs6$ \\\hline
\end{tabular}
\end{minipage}
\end{table*}

As RM Synthesis requires a Fourier inversion of a windowed
function, the Faraday dispersion function, $F(\phi)$, is convolved with a point-spread function in Faraday space. This point-spread function is known as the
Rotation Measure Spread Function (RMSF), and has a FWHM given by
\begin{equation}
  \label{RMSF-FWHM}
  \delta\phi \approx \frac{ 2 \sqrt{3} }{\Delta\lambda^2} \,,
\end{equation}
where $\Delta\lambda^2$ is the width of the $\lambda^2$ distribution
i.e.\ the range in $\lambda^2$ of the observing bandwidth. However, the precision in a measurement of the peak Faraday depth is a function of the FWHM of the RMSF divided by twice the signal--to--noise ratio (s/n) \citep{2005A&A...441.1217B}. In principle, the RMSF can be deconvolved
from $F(\phi)$ using a one-dimensional clean that is analytically
identical to that used during aperture synthesis imaging
\citep{2009A&A...503..409H}.
%
%
%
%
There are clearly many analogies between RM
Synthesis and aperture synthesis, with the concepts of
$\lambda^2$-coverage and $uv$-coverage, and of an RMSF and a synthesised
beam \citep[see][for more detail]{2005A&A...441.1217B}. RM Synthesis
therefore provides a useful tool with which to probe magnetic fields in
the Universe.

\section{GMRT Observations and Data Reduction}\label{S:GMRT}

The GMRT observations are summarised in Table~\ref{tab:obs}. Both
observations used the flux calibrators 3C138 and 3C48. Observation
17\_060\_1 used the phase calibrator J0240$-$231, while 17\_060\_2 used
J2314$-$449. The derived fluxes were found to be $5.23\pm0.06$~Jy and
$3.05\pm0.04$~Jy respectively. Data reduction and calibration were
carried out using standard tasks in the 31DEC10 AIPS package.

All GMRT observations were taken in spectral-line mode, which assists
with excising narrow-band interference. Radio-frequency interference
(RFI), pointing errors, and problems with position control (i.e.\ the
servo system) result in anomalous data being recorded by an
interferometer. These bad data need to be removed (`flagged') from the
observation. Full-polarisation data taken with the GMRT require
particularly extensive flagging, as the visibilities are dominated by
RFI which frequently impinges on the band. Bad data were identified
using the tasks \textsc{tvflg} and \textsc{spflg}. The data were first
flagged in the $RR$ and $LL$ cross-correlations, and the entire
procedure then repeated for $RL$ and $LR$. In principle, the weaker
signal in $RL$/$LR$ should allow flagging to be accomplished using these
cross-correlations alone. However, checks on the data indicate that RFI
remains in the $RR$ and $LL$ visibilities following flagging in
$RL$/$LR$. All cross-correlations were therefore flagged individually.
As the polarisation calibration is performed on each single channel,
low-level RFI can result in serious errors and a poor calibration. The
only way found to be effective for manually flagging full-polarisation
GMRT data was to ensure that every baseline and channel was checked in
$RR$, $LL$, $RL$, and $LR$.

Corrections were then made for antenna-based bandpass effects, the flux
density scale, and amplitude and phase calibration using well-documented
techniques \citep[e.g.][]{2009PhDT.........3G, mythesis}. The flux scale
was determined using the `Perley--Butler 2010' coefficients. Channels at
the edges of the band were excluded from the calibration, resulting in
220 usable channels. The GMRT has a maximum baseline length of $25$~km, providing a typical resolution of $\approx5$~arcsec at $610$~MHz. Due to the loss of outer arm antennas and gaps in the $uv$-coverage, we frequently smooth the data to a larger FWHM during the imaging process. Where appropriate, the FWHM of the restoring beam is shown in each Figure.

\subsection{Instrumental Leakage}
\label{onaxisleakagesection}
\begin{figure*}
\centering
\includegraphics[clip=true,trim=0cm 0cm 0cm 0cm,width=15cm]{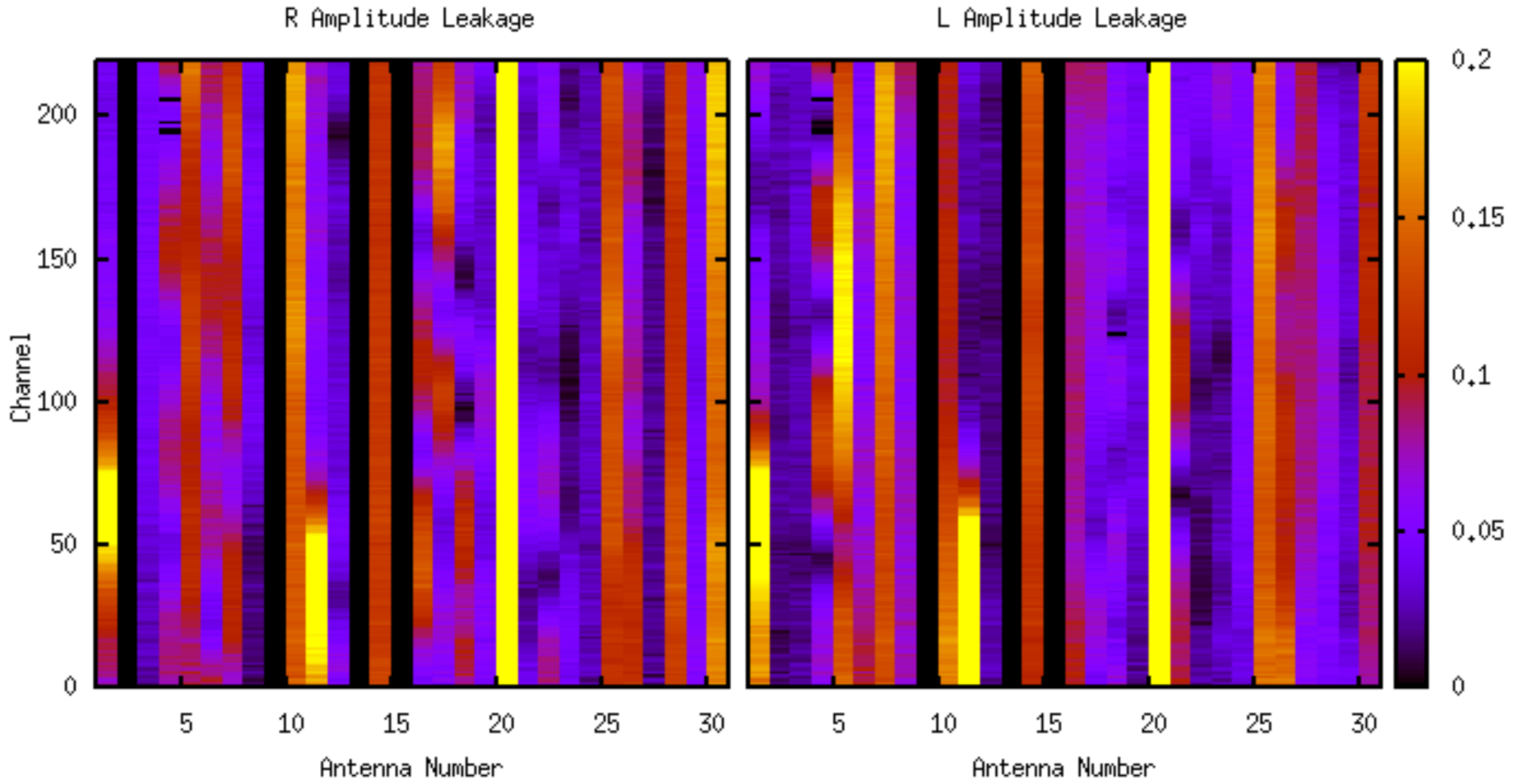}
\caption{The amplitude of the leakage terms $d_{jR}$, $d_{jL}$ (see Section~\ref{onaxisleakagesection} for further details) for observation 17\_060\_2, in both $R$ (left) and $L$ (right). The $x$-axis shows the antenna number, the
$y$-axis shows the channel number across the observing bandwidth, and
the colour-scale shows the leakage amplitude. Antennas coloured black
across the entire band are flagged.}
\label{leakages}
\end{figure*}

Following Hamaker et al. (1996), the response of an interferometer can be described by
antenna-based response matrices, known as Jones matrices. The response of antenna $j$ with
orthogonal circularly polarised feeds ($R$ and $L$) can be expressed in terms of a Jones matrix
that operates on the column vector $(R, L)^T$. The multiplication order is the inverted sequence of effects that take place as the signal propagates from the source to the antenna. The Jones matrix is then given by
\begin{equation}
J_{j} = G_{j} D_{j} P_{j} \,,
\end{equation}
where $G_{j}$ is the gain, with the matrix representing the effects of the atmosphere and the electronics, and $P_{j}$ describes the effects of the rotation of an altitude--azimuth (alt--az) mounted antenna as seen by a tracked source. The `$D$-term' is given by
\begin{equation}
D_{j} = \left( \begin{array}{cc}
1 & d_{jR} \\
-d_{jL} & 1  \end{array} \right) \,,
\end{equation}
which models the deviations of the feed's polarisation response from that of an idealised system. The complex leakage terms $d_{jR}$, $d_{jL}$ represent the fraction of the orthogonally polarised signal `leaking' into a given feed. In practice, antenna feeds are never perfectly circularly polarised. This leads to detectors sensitive to $R$ also detecting a certain percentage of $L$ and vice-versa. These imperfections can be modelled by a `polarisation leakage'. This leakage, or `instrumental polarisation', was found to be
highly frequency-dependent at the GMRT \citep[also see][]{2009ASPC..407...12T}. The
typical leakage amplitude is $\approx5{-}10$\%, with a few antennas
having leakages of up to $\sim40$\%, as shown in Fig.~\ref{leakages}.
The instrumental polarisation must therefore be calculated for each
individual spectral channel. Separation of instrumental and source
contributions was achieved through repeated observation of the phase
calibrators (listed at the start of this section) over a large range in parallactic angle, $\chi$. Such a technique is advantageous in that the source can be polarised or unpolarised, for further details see e.g.\ \citet{1994ApJ...427..718R}. This was done with the AIPS task \textsc{pcal} using a linearised model of the feeds
(\textsc{soltype=`appr'}). The data were time-averaged before
calibration into 2 minute intervals (\textsc{solint=2}). Varying the
time-averaging between 1--10 minutes was found to have a negligible
effect on the polarisation calibration. As the linearised model loses
accuracy for large leakages, the quality of calibration was limited for
several antennas in the array with high leakage amplitudes.
Consequently, all antennas with leakages $\ge$15\% were flagged from the
data. This removed 7 antennas from each observation. There is an
associated loss in sensitivity, and occasionally outer arm antennas had
to be flagged -- leaving relatively large gaps in the $uv$-coverage. An
appropriate selection of $uv$-range and/or $uv$-taper was used during
the imaging stage to accommodate for these gaps. Extended polarised
emission remains well-sampled by the over-density of antennas in the
GMRT's central square. The minimum baseline length of $\sim 100$~m means
the data is sensitive to polarised emission on scales up to $\sim
17$~arcmin. Checks on the calibrators 3C48 and 3C138 show that the
residual instrumental polarisation is $\le0.25$\%, providing a check on both an unpolarised and polarised source at these frequencies \citep{2013ApJS..206...16P}.

\subsection{Ionospheric Faraday rotation}

Prior to instrumental polarisation calibration, small corrections for
the effect of ionospheric Faraday rotation were made using the task
\textsc{tecor}. These corrections were enacted together with an
ionospheric model produced by the Jet Propulsion Laboratory, that is
freely available from the CDDIS data archive. The maps were used to
correct for ionospheric Faraday rotation at the reference antenna, with
differences between the antennas being removed by self-calibration. The
GPS measurements of the vertical total electron content (TEC) are
spatially interpolated and temporally smoothed, and updated every 15
minutes. The nearest GPS receivers to the GMRT are located in Bangalore
and Hyderabad. There is currently no widely available model to correct
for differences in the observing geometry or the ionosphere's magnetic
field strength. Nevertheless, the magnetic field is only weakly variable
over the duration of a typical observation, while the TEC can be highly
variable (Cotton 1993). Maps of the TEC are therefore an important
first-order correction, and were found to reduce the variance of the
$RL$/$LR$ phase in the GMRT data from $18^\circ$ to $15^\circ$, and
typically by $\approx 1^\circ{-}3^\circ$. The maximum Faraday rotation
due to the ionosphere was estimated by the model to be $\le
2$~rad~m$^{-2}$ over the duration of each observation.

\subsection{Time-Stability of the Leakages}

Due to s/n constraints, it is generally only possible
to calculate a \emph{single} leakage for each antenna in a
full-synthesis observation. However, each observation actually takes
place over many hours. In order to calibrate the data, we must therefore
make the critical assumption that the antenna's on-axis response is
sufficiently stable that it can be parameterised by a single complex
leakage for the entirety of an observation.

Time-dependent leakages would result in uncorrected `residual'
instrumental polarisation. This residual polarisation would cause
emission in Stokes $I$ to leak into and corrupt $Q$ and $U$.

The short-term time-stability of the leakages was investigated by
initially solving for the instrumental and source polarisation
simultaneously -- using the phase calibrator over the full range in
parallactic angle. The source polarisation determined via this process
was then used as an input model for future calculations of the
instrumental polarisation. This allowed for the instrumental
polarisation to be calculated in a manner similar to when a source of
known polarisation is used.

This method allowed the leakages to be calculated for each individual
$\sim5$~minute scan of the phase calibrator throughout four consecutive
observations, two of which were 17\_060\_1 and 17\_060\_2. This yielded
49 calculations of the instrumental leakage in each spectral channel
during the two day observing period. The RMS variation in these leakages
will be a consequence of two effects: a contribution due to thermal
noise, and an additional component due to time-variability of the
leakages. An analysis of the time-variability is therefore reliant on
the ability to separate out these two contributions. As the observations
use four separate phase calibrators, the data also allow for any
significant elevation-dependence to be identified.

\begin{figure*}
\centering
\includegraphics[clip=true,trim=0cm 0cm 0cm 0.5cm, width=14cm]{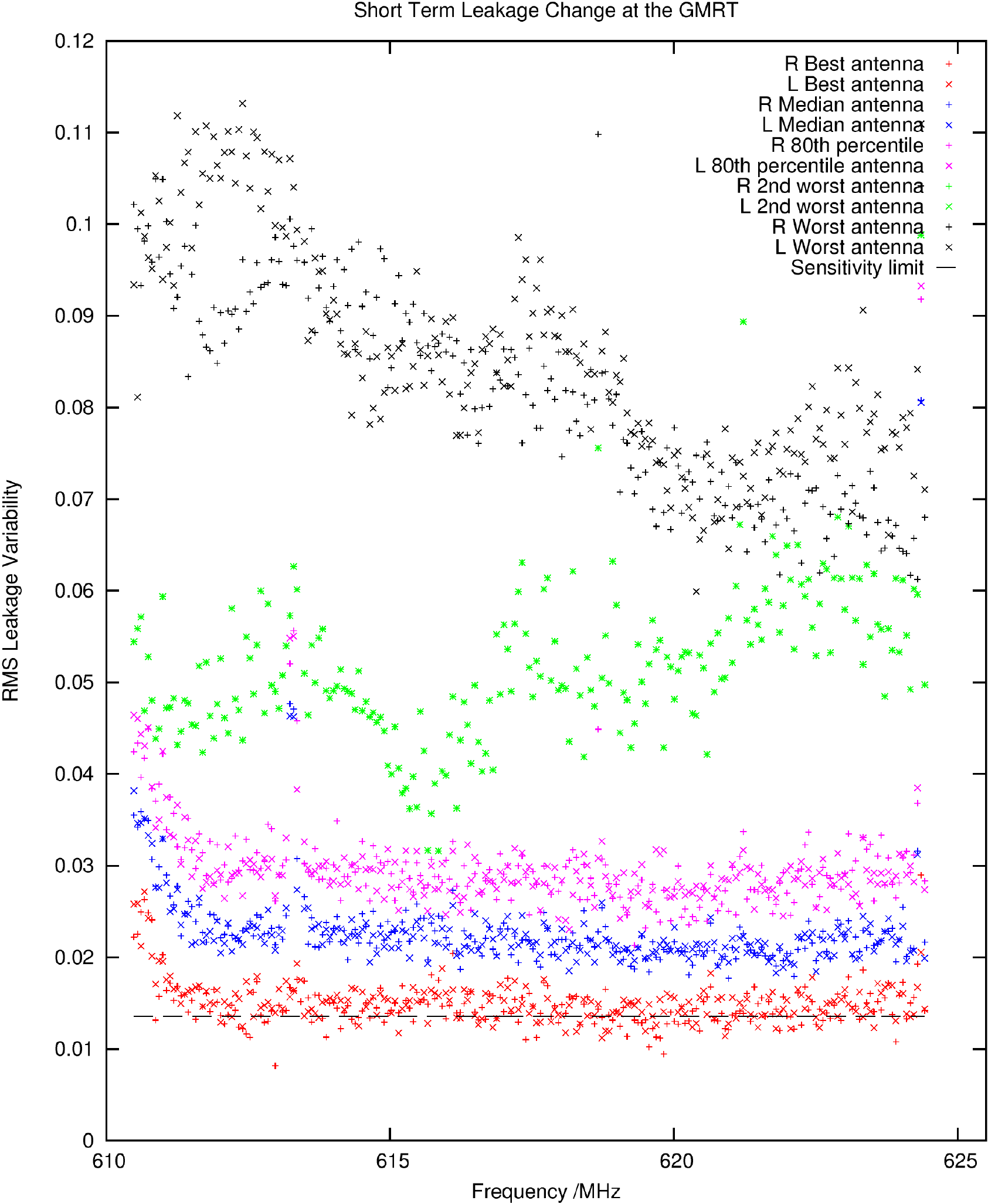}
\caption{The short-term time-variability of instrumental polarisation
across the GMRT band at 610~MHz. The `RMS leakage variability' has been
calculated about the mean of the derived solutions. The best, median,
80th percentile, 2nd worst, and worst antenna, in terms of their RMS
variability are shown. Both $R$ and $L$ are plotted. A line showing the
sensitivity limit of the four observations (i.e.\ $\sigma_{D}$) is also displayed and shows
the expected RMS variation due to thermal noise; this was found to be
approximately constant across the band.
}\label{Variability}
\end{figure*}

The methodology used by \citet{EVLAMemo} was adopted, and they find that
the uncertainty due to system noise is approximately given by
\begin{equation}
  \label{leakagenoise}
  \sigma_{D}^2 = \frac{1}{N} \frac{\sigma_{C}^2}{I^2} \,,
\end{equation}
where $\sigma_{D}^2$ is the variance of the real or imaginary parts of
the leakages, $\sigma_{C}^2$ is the variance of the correlation data,
and $N$ is the number of antennas. $\sigma_{C}^2/I^2$ can be estimated
from the RMS of the closure phase, which is expected to be equal to
$3\sigma_{C}^2/I^2$. To calculate $\sigma_{D}$, the RMS of the closure phase was extracted from the data using the AIPS task \textsc{shouv}, and was
found to be approximately constant across the band. Equation
\ref{leakagenoise} is derived using the approximation that the error in
each leakage term is independent. This is incorrect as the leakages are
derived per antenna and not per baseline, but this only affects the
results by a factor of order $1/N$. We find $\sigma_{D}\approx0.0135$. A plot showing the RMS leakage variability across the band is shown in
Fig.~\ref{Variability}.

The variation between the best, median, and 80th percentile antennas is approximately twice that of the RMS variability expected due to thermal noise. This range is not large, it is only 2.2$\sigma$. However, there are a number of anomalous antennas which seem to have
significant time-dependent leakage. Across the entire band, the worst
antenna was C09 -- with an RMS leakage variability $\approx8$ times
larger than that expected due to thermal noise. Antennas C06 and E02
also possibly had statistically significant time-dependent leakage, with
a variability $\approx4$ times that expected due to noise. The effect of
C06 and E02 on the interferometric images was investigated. C06 appeared
to be well-calibrated, and removal of the antenna had the sole effect of
increasing the noise in the subsequent images. This leads us to conclude
that antenna C06 was simply a noisy outlier during these observations.
Nevertheless, the removal of E02 visibly reduced image artefacts
surrounding polarised sources.

After polarisation calibration, the matrix of $RL$ and $LR$ phases
displayed by \textsc{rldif} showed that the mean phase of antennas C09
and E02 was offset from the matrix average. The offset was present in
all four of the observations used for the analysis. While C09 has the
highest leakage of all antennas during these observations, E02 has
leakage below the mean level. The data appear consistent with the
effects of residual time-dependent instrumental polarisation. The cause
of this time-dependent instrumental polarisation at 610~MHz is unknown.
Both antennas were flagged -- with the $RL$ and $LR$ visibilities being
removed from every observation. Both antennas are physically located
within or near to the central square, which fortunately minimises the
size of the gaps in the $uv$-coverage that results from antenna removal.
The issue with these antennas was provided to the team at the
observatory. Following maintenance, this offset was not present for
either antenna in follow-up observations taken in 2011 January.

Taking into account the sensitivity limit of these observations, the
instrumental polarisation of the median antenna is essentially
independent of time. Leakages vary by $<2$\% over a timescale of a few
days, and $\lesssim0.5$\% over the length of a typical observation.

\subsection{Polarisation angle calibration}

When solving for the instrumental leakage, the absolute value of the
phase offset between $R$ and $L$ is left as an unconstrained degree of
freedom. A change in this $R-L$ phase difference is equivalent to a
change of parallactic angle, therefore causing a rotation of $Q \pm iU$
-- which is also equivalent to a change in the electric vector
polarisation angle (EVPA) \citep[e.g.][]{HamakerREF}.

Following leakage calculation at the GMRT, the variation in the EVPA
across the band conspires such that every source has an `instrumental
RM' with a typical magnitude of several hundred rad~m$^{-2}$. For 3C138
in observation 17\_060\_2, the instrumental RM$=-678\pm5$~rad~m$^{-2}$.
This instrumental RM was corrected for each spectral channel using
3C138, which was taken to have an RM of $4.0$~rad~m$^{-2}$
\citep{mythesis}.

The typical method of using the average phase from the matrix of the
$RL$/$LR$ correlations was found to be noisy and to provide inconsistent
results. We instead calibrated the EVPA using the integrated flux of
3C138 as determined from the Stokes $Q$ and $U$ images, with the
appropriate correction factor being applied to the $uv$-data. No
significant emission was found in Stokes $V$.

\subsection{Direction-dependent instrumental polarisation}
\label{offaxispol}
The response of an interferometer varies across the primary beam, and
wide-field polarimetry requires calibration of these
`direction-dependent' or `off-axis' instrumental effects. Similar to the
case of on-axis calibration, the `polarisation beam' manifests itself
with flux in total intensity leaking into the polarisation images. These
polarimetric aberrations result in an increase in the observed
fractional polarisation and also alter the absolute EVPA of sources --
with the effect becoming more pronounced with increasing distance from
the phase-centre. Direction-dependent effects therefore limit the
dynamic range of low-frequency interferometric images, and restrict the
region of the primary beam that is useful for scientific measurement.

The general equations to compute the off-axis instrumental polarisation beam are given by \citet{caretti2004}. From a more observational perspective, and following \citet{HeilesAreciboMemo}, there are two kinds of beam
polarisation that are theoretically expected:
\begin{enumerate}
\item Beam squint: this occurs when the two circular polarisations point
in different directions by a certain angle. Beam squint tends to produce
a `two-lobed' pattern, with the beam having one positive and one
negative region.
\item Beam squash: this occurs when the beamwidths of the two linear
polarisations differ by a certain amount. Beam squash tends to produce a
`four-lobed' pattern, in which two regions on opposite sides of the beam
centre have a positive sign and two lobes rotated by $90^{\circ}$ have a
negative sign. This quadrupolar pattern tends to give rise to
instrumental linear polarisation that is oriented radially with respect
to the phase-centre.
\end{enumerate}

We measured the GMRT polarisation beam by taking complementary
observations in modified holography mode. Two reference antennas
remained fixed on an unpolarised calibrator (3C147) while the remaining
antennas were slewed in azimuth and/or elevation to a co-ordinate such
that the calibrator source was observed offset from the phase-centre. In
all cases, the two reference antennas were excluded from the data
analysis so that only the offset antennas were used -- allowing the
power pattern to be directly retrieved, rather than the voltage pattern.

\begin{figure}
\centering
\includegraphics[clip=true,trim=0cm 0cm 0cm 0cm,width=8.4cm]{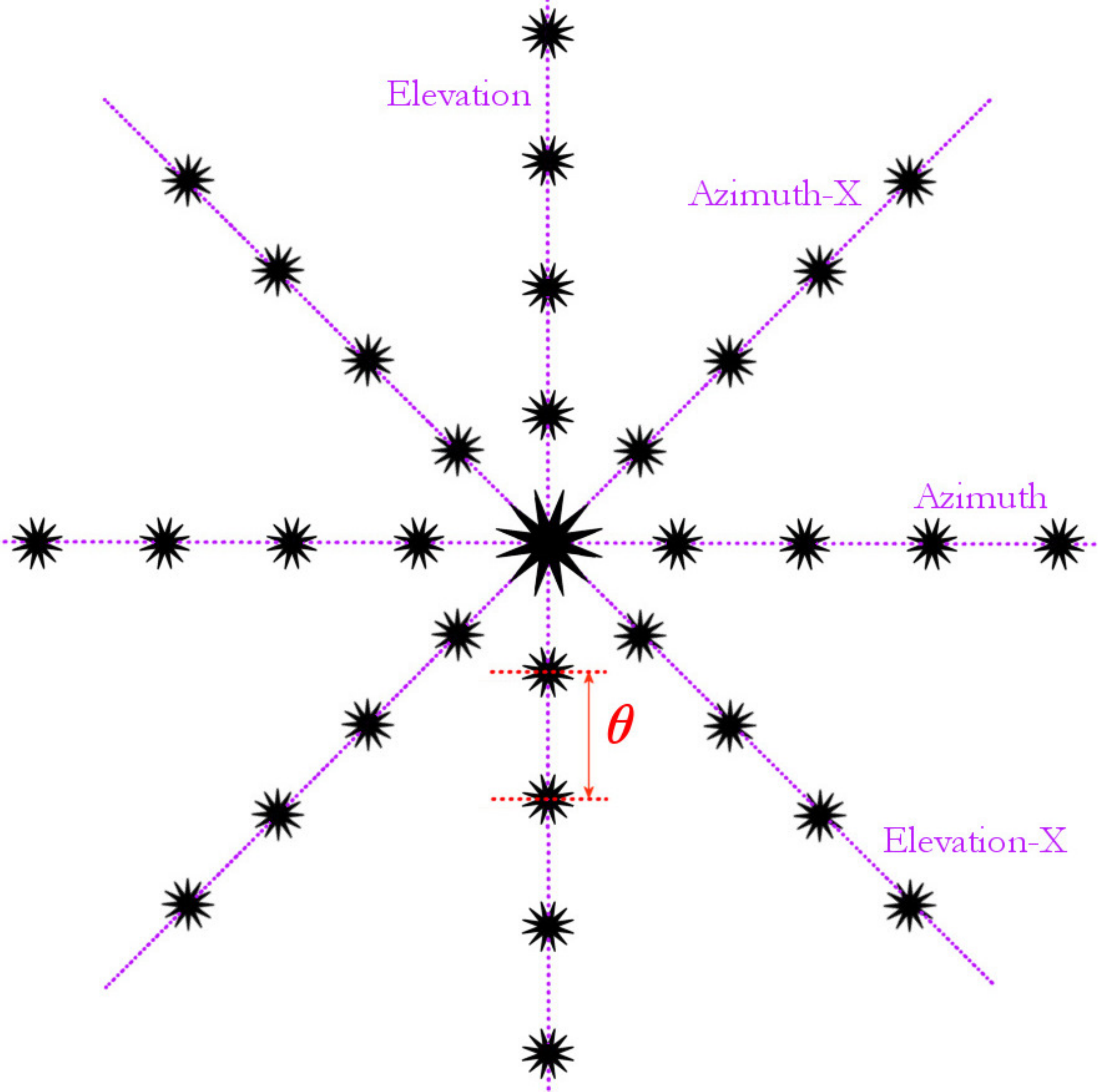}
\caption{The grid used for observing the off-axis raster scans in order to estimate the polarisation beam response. Each
axis is labelled on the end that is defined as positive. For these
observations, $\theta = 10$~arcmin, with 9 raster scans along each axis
and a maximum offset of $40$~arcmin. The FWHM of the primary beam, as
measured at the observatory is $44\farcm4$ at
$610$~MHz.
}\label{offaxisgrid}
\end{figure}

A number of offsets were used, such that a grid of `raster scans' were
observed. The grid used for observing these off-axis raster scans is
shown in Fig.~\ref{offaxisgrid}. An EVPA calibration was applied to
obtain the $Q$ and $U$ beam in the calibrated antenna frame. This
correction to the holography data was made by assuming that the EVPA of the azimuth axis was oriented radially outwards from the phase-centre.

\begin{figure*}
\centering
\includegraphics[clip=true, trim=0cm 0cm 0cm 0cm, width=12cm]{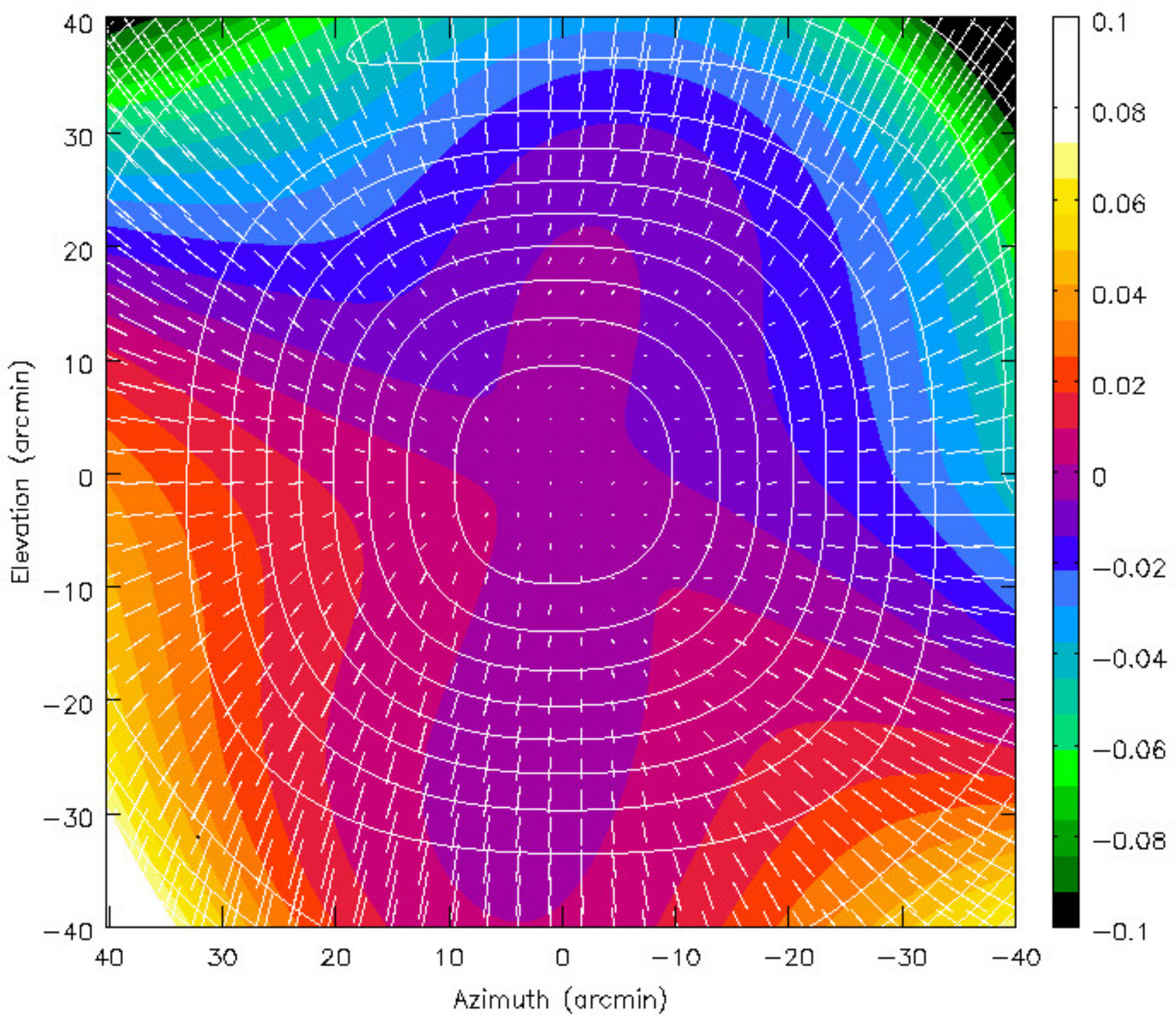}
\caption{The main lobe of the instantaneous GMRT primary beam at 610~MHz. Contours show the Stokes $I$ response with contour levels at 10, 20,
30\ldots90\%. Polarisation vectors are proportional to the
linearly polarised intensity and indicate the orientation of the
linearly polarised response. At the 50\% Stokes $I$ contour, the polarisation vectors show a typical leakage of $\sim10$\%. The pseudocolour scale shows the instrumental circular polarisation from $-10$\% to $+10$\%. There are artefacts at the outer periphery
of the beam that result from interpolating over a greater tangential
distance at large radii.
}\label{allbeam}
\end{figure*}

\begin{figure*}
\centering
\includegraphics[clip=true, trim=0cm 0cm 0cm 0cm, width=8.5cm]{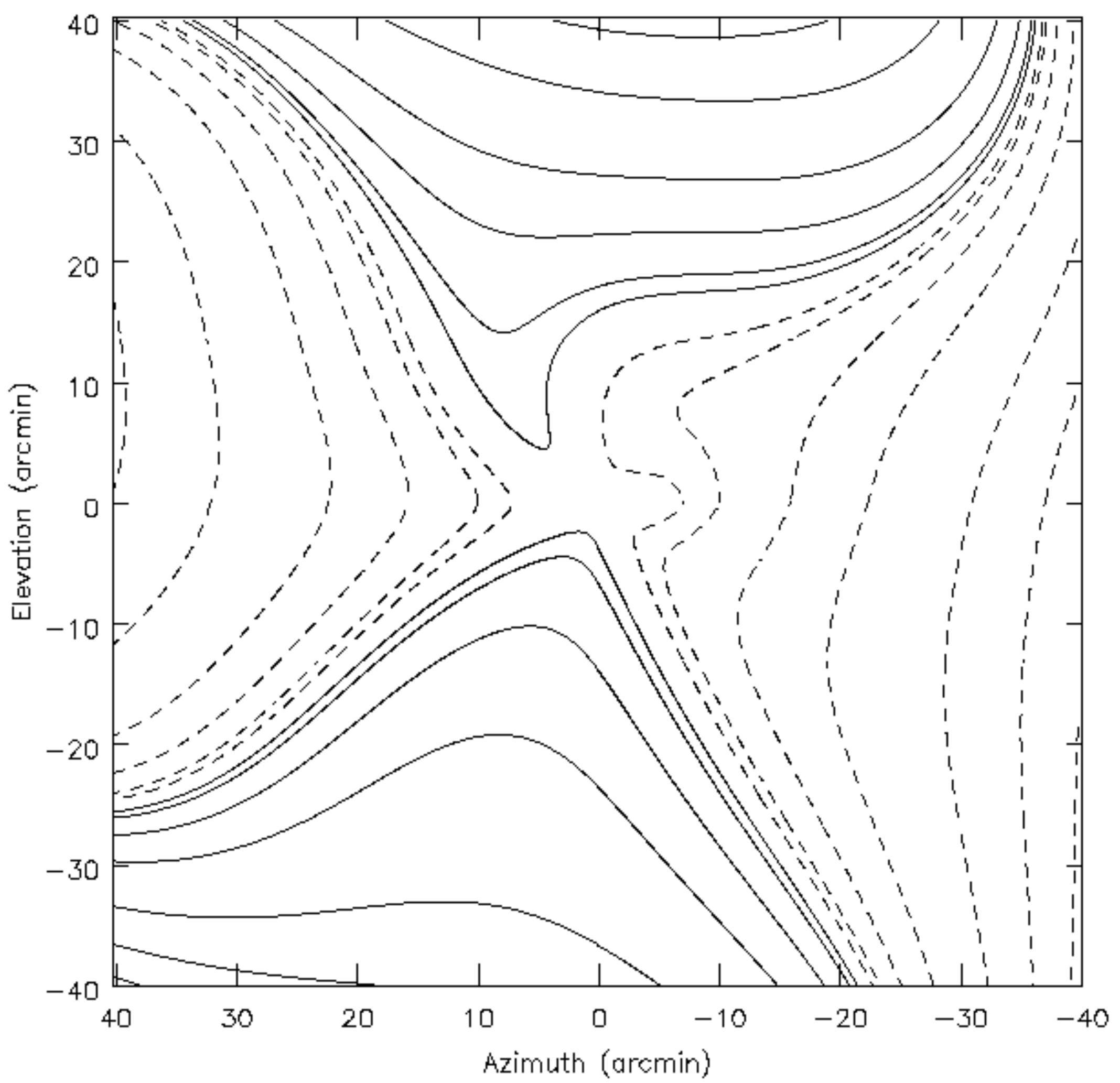}
\includegraphics[clip=true, trim=0cm 0cm 0cm 0cm, width=8.5cm]{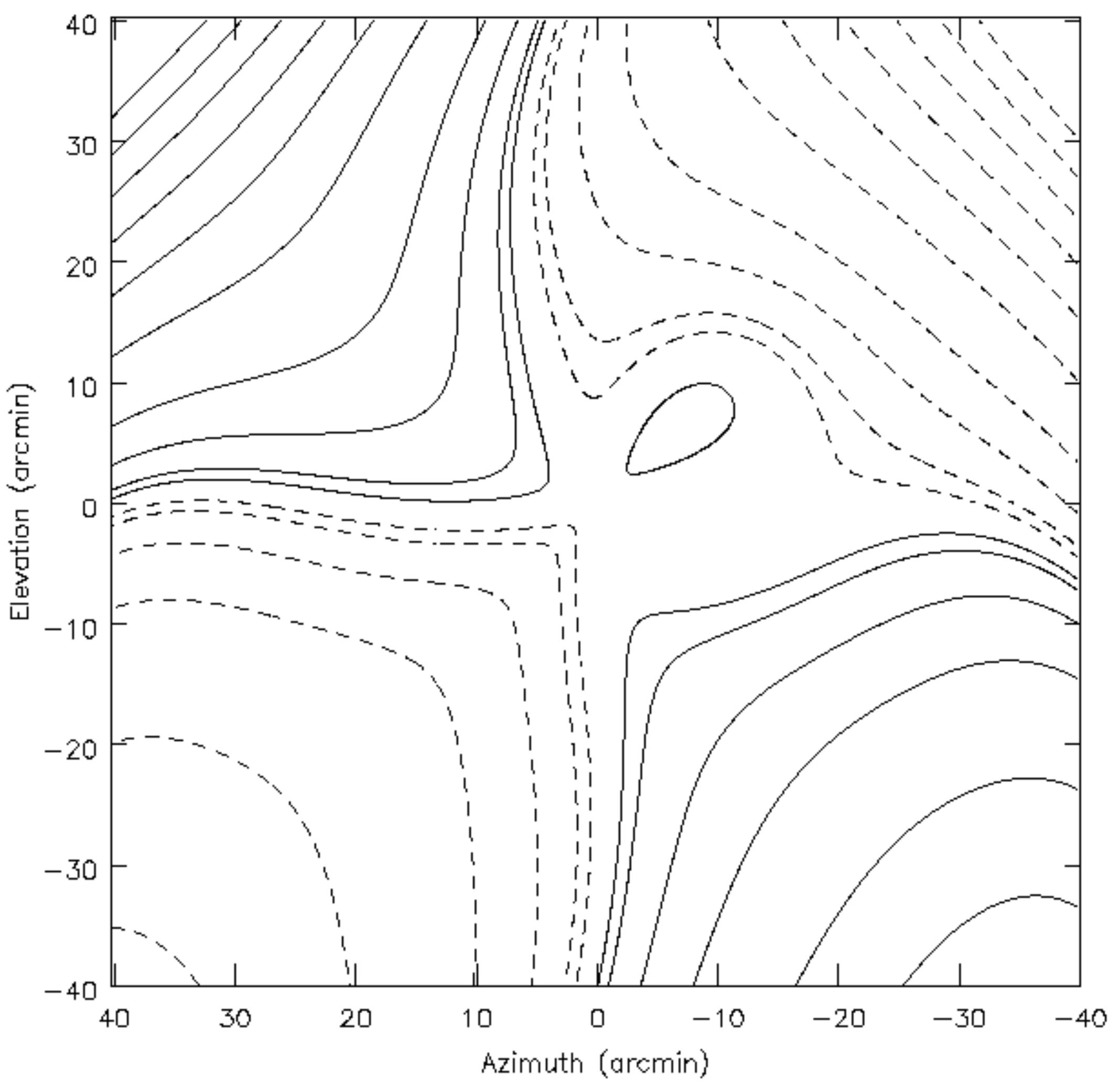}
\caption{The Stokes $Q$ (left) and $U$ (right) instantaneous fractional instrumental polarisation beams of the main lobe of the GMRT at 610~MHz. Contour levels are at $\pm$1, 2, 5, 10, 20, 30\ldots70\%.
}\label{QUbeam}
\end{figure*}

By outputting the average amplitude and phase of all visibilities for
each holography raster in all four cross-correlations, it is possible to
parameterise the direction-dependent response in terms of the response
of the \emph{average} antenna. Variations between antennas have not been
considered here. For an ideal, calibrated interferometer with circular
feeds, the cross-correlations are complex quantities defined as
\begin{equation}
  RR = \mathcal{A}(RR)\textrm{e}^{i\psi_{RR}} = I+V \,,
  \label{RRcomplex}
\end{equation}
\begin{equation}
  LL = \mathcal{A}(LL)\textrm{e}^{i\psi_{LL}} = I-V \,,
  \label{LLcomplex}
\end{equation}
\begin{equation}
  RL = \mathcal{A}(RL)\textrm{e}^{i\psi_{RL}} = Q+iU \,,
  \label{RLcomplex}
\end{equation}
\begin{equation}
  LR = \mathcal{A}(LR)\textrm{e}^{i\psi_{LR}} = Q-iU \,,
  \label{LRcomplex}
\end{equation}
where $\mathcal{A}(jk)$ and $\psi_{jk}$ are the amplitude and phase
respectively. For a calibrated point-source, $\psi_{RR}=\psi_{LL}=0$, so
the phase terms of equations \ref{RRcomplex} and \ref{LLcomplex} can be
neglected. However, the terms $\psi_{RL}$ and $\psi_{LR}$ cause mixing
of $Q$ and $U$ and cannot be ignored.

By applying Euler's formula to equations \ref{RLcomplex} and
\ref{LRcomplex}, and substituting into $RL=LR^{\ast}$, the solutions can
be used to  obtain the fractional polarimetric beam response at a given
raster. This was done for every raster in each of the individual 220
channels across the band. The holography source was taken to be
unpolarised, so no correction for source polarisation was required.

The instrumental polarisation and effects of parallactic angle ($\chi$) rotation have only
been corrected at the phase-centre. These effects are actually
direction-dependent and due to the GMRT's alt--az mount, any residual
instrumental polarisation will display an EVPA that rotates with $\chi$. All off-axis holographic raster scans had to
be corrected for the effects of this $\chi$-dependent mixing of $Q$ and
$U$, so that
\begin{equation}
  (Q' + iU') = (Q + iU)\textrm{e}^{-2i\chi} \,.
  \label{QUmixing}
\end{equation}
The derotated $Q'$ and $U'$ can then be used to form a beam profile in
full-Stokes for each channel across the band. The beam was found to be independent of frequency across the observing bandwidth. The mean response across
the band was obtained for each holography raster, so that the beam was
described by a single datum per raster -- creating a set of nine data
per beam axis. Each axis was then fitted with a 5th order polynomial in
Stokes $I$, $Q$, $U$, and $V$ using an ordinary least-squares fit. The
polynomial fits were then used to create a two-dimensional map of the
GMRT beam at 610~MHz by tangentially interpolating and gridding the beam
response around the phase-centre at a given radius. This was carried out
using cubic spline interpolation in order to smoothly interpolate the
sparsely sampled data over the large tangential distances.

The polarisation beam response along the diagonal axes suggested a
significant problem with the Elevation-X axis. The direction-dependent
response along this axis was observed to be oriented in the same
direction as the Azimuth-X axis -- in contradiction to the expected,
previously measured quadrupolar response \citep{2009MNRAS.399..181P,
FarnesRMAA}. Further investigation of our data showed that an
observational bug led to the re-observation of the Azimuth-X axis instead
of independently measuring the Elevation-X axis. We therefore lack data
from the true Elevation-X axis. The orientation of polarisation vectors
along the observed Elevation-X axis were therefore rotated so that they
were oriented radially -- this rotation was equivalent to a change of
sign for both $Q$ and $U$. The resulting maps therefore use the
Azimuth-X axis twice, under the assumption that the true Elevation-X
axis is similar to the Azimuth-X, but maintains a radial orientation.
This assumption is justified as the resulting maps are consistent with
the response of the GMRT inferred using both RM Synthesis
\citep{FarnesRMAA}, and from antenna theory \citep{2009MNRAS.399..181P}.
The final two-dimensional beam at 610~MHz is shown in
Fig.~\ref{allbeam} and \ref{QUbeam}. It is possible to correct for the effects of wide-field instrumental polarisation by calibrating the $uv$-data using the beam model for each `time-chunk' of approximately constant parallactic angle \citep[e.g.][]{mythesis}. Nevertheless, the process is highly computationally expensive. These observations have therefore not been corrected for these wide-field instrumental effects. However, our constraints on the polarisation beam -- particularly the radial orientation of the polarisation vectors, which averages down the instrumental effect in our full-track observations -- allow us to be confident in our polarisation detections across the field of view. Where appropriate, we have provided an estimate of the residual instrumental effects alongside each of our off-axis polarisation measurements.

\subsection{Imaging and RM Synthesis}

Following calibration, each channel was imaged in Stokes $Q$/$U$/$V$
individually using multi-facet imaging -- breaking the sky up into 31
facets across the field of view (FOV). The resulting Stokes $Q$ and $U$
maps were combined into images of polarised intensity, $P = \sqrt{Q^2 +
U^2}$, and the resulting 220 images of $P$ were averaged together into a
final image of the band-averaged polarised intensity. As the images of
$P$ are averaged, rather than averaging $Q$/$U$, the images are not
subject to bandwidth depolarisation -- and therefore provide a useful
diagnostic of the polarisation across the FOV. Nevertheless, full RM
Synthesis is needed to retrieve the properties of the polarised
emission.

\begin{figure}
\centering
\includegraphics[clip=true, trim=0.2cm 0cm 0.1cm 0cm, width=8.4cm]{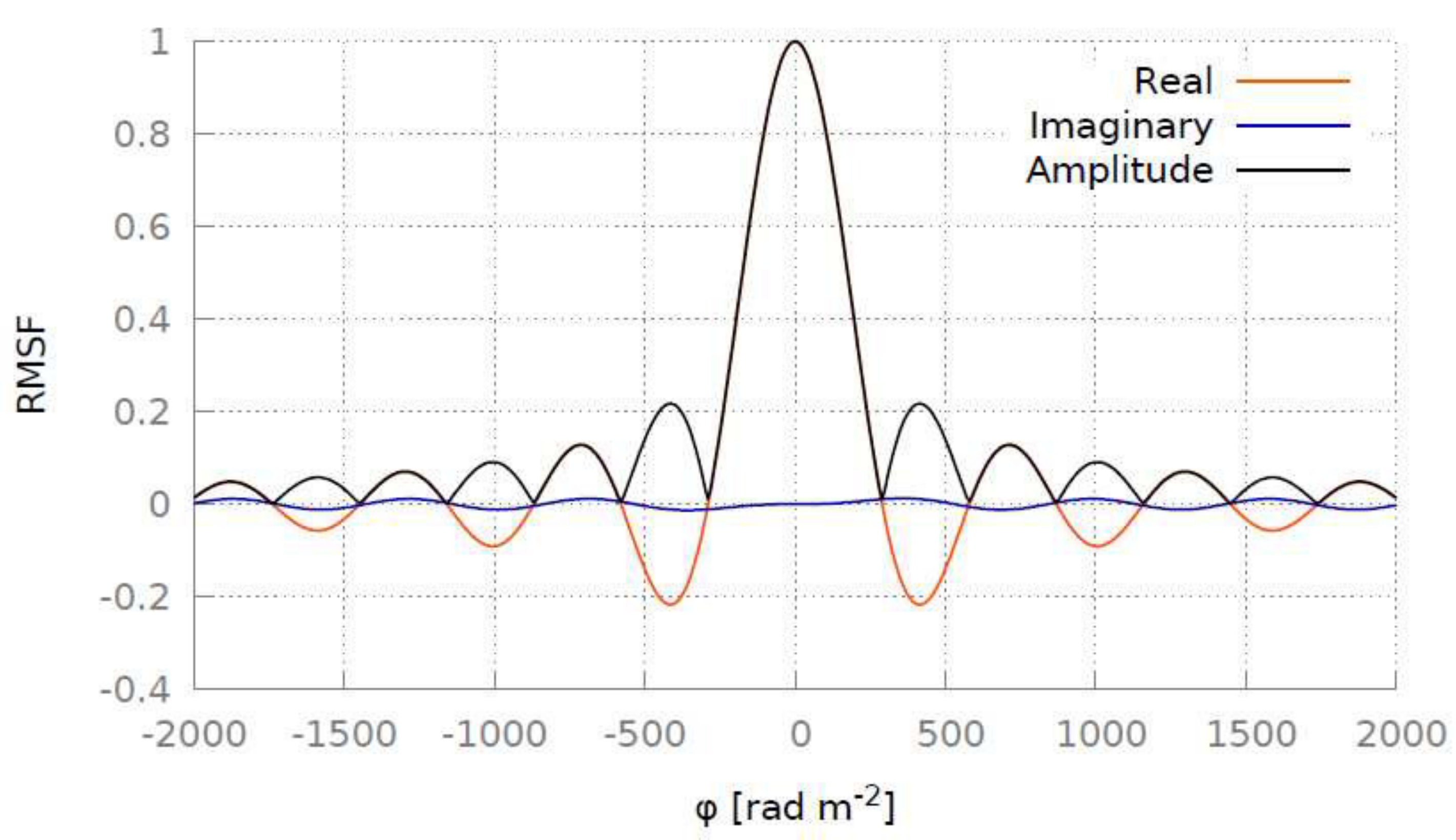}
\caption{The rotation measure spread function (RMSF) for these GMRT
observations. The real and imaginary components ($Q+iU$) and the
amplitude ($P = \sqrt{Q^2 + U^2}$) are shown.
}\label{RMSF}
\end{figure}

The Stokes $Q$ and $U$ images were therefore used to form a
three-dimensional datacube, so that the technique of RM Synthesis could
be applied \citep{2005A&A...441.1217B}, using a code developed in
Python. This code also implemented a form of RM-clean
\citep{2009A&A...503..409H} to deconvolve the Faraday dispersion
function (FDF) from the Rotation Measure Spread Function (RMSF). The
version of RM-clean used here considers the effects of the residuals,
and uses complex cross-correlation in order to locate the peak in
Faraday space \citep{2009A&A...503..409H}. There is currently some
question over the capability of RM-clean to fully reconstruct the
Faraday dispersion function along the line of sight
\citep{2010MNRAS.401L..24F, 2011AJ....141..191F}, and it is possible
that modelling of $Q(\lambda^2)$ and $U(\lambda^2)$ is advantageous
under many circumstances \citep{2012MNRAS.421.3300O}. These issues are
not the limiting factor for our data, due to the 16~MHz bandwidth and
the corresponding FWHM of the RMSF = 321~rad~m$^{-2}$. The RMSF is shown
in Fig.~\ref{RMSF}. RM-cleaning was performed down to a $2\sigma$ limit.
The cleaned Faraday spectrum was convolved with a FWHM equal to that of
the RMSF. Unless otherwise specified, the polarisation was summed up
coherently across the band for trial Faraday depths ranging from
$-2000$~rad~m$^{-2}$ to $+2000$~rad~m$^{-2}$. The `RM-cubes' generated
by RM Synthesis were always oversampled -- sampling of $\Delta\phi =
1$~rad~m$^{-2}$ was often used. The RM Synthesis observational setup,
which uses a 16~MHz bandwidth, with 62.5~kHz channel spacing at 610~MHz,
has a maximum Faraday depth to which one is sensitive of $|\phi_{\textrm{max}}|
\sim 35000$~rad~m$^{-2}$, and a FWHM of the RMSF of 321~rad~m$^{-2}$. No
attempt was made to distinguish multiple Faraday components along a
single line of sight. Bandwidth depolarisation will affect these data
for a $|$RM$|\ge20000$~rad~m$^{-2}$. The maximum scale in Faraday space
to which the data are sensitive is 14~rad~m$^{-2}$, consequently the
data are not sensitive to Faraday thick emission. Faraday depths (FDs)
were retrieved using a least-squares Gaussian fit to the peak of the
deconvolved FDF.

\subsection{Spectral Indices}

\begin{table*}
\begin{minipage}{110mm}
\caption{Parameters of the observed SCGs.}\label{tab:groupinfo}
\begin{tabular}{@{}ccccccc@{}} \hline
Group & $D^{\rm a}$ & $N_{\rm Gal}$ & Galaxy Name & Label & Velocity &
   Type$^{\rm c}$ \\
 & /Mpc &  &  &  & /km~s$^{-1}$ & \\\hline
Grus Quartet & 22.5 & 4 & NGC~7552 & $a$ & 1611 & SB(s)ab \\
 & & & NGC~7582 & $b$ & 1575 & SB(s)ab, Sy2 \\
 & & & NGC~7590 & $c$ & 1578 & S(r?)bc \\
 & & & NGC~7599 & $d$ & 1653 & SB(s)c \\ \hline
USCG~S063 & 54.5 & 5 & IC~1724 & $A$ & 3804 & S0$^{+}$ \\
 & & & IC~1722 & $B$ & 4131 & SAB(s)bc \\
 & & & ESO~0353$-$G036 & $C$ & 3822 & SB(rs)0, SFG \\
 & & & IRAS~F01415$-$3433 & $D$ & 3759 & S0 \\
 & & & ESO~0353$-$G039 & $E$ & 3915 & Sc, SFG \\\hline
\end{tabular}

Notes:\\
a) The co-moving distance, assuming a flat cosmology with $\Omega_{M} =
0.27$, $\Omega_{\Lambda} = 0.73$, and $H_{0} =
71$~km~s$^{-1}$~Mpc$^{-1}$.\\
b) Calculated using $v=cz$ with $z$ taken from the NASA/IPAC
Extragalactic Database (NED) (\texttt{http://ned.ipac.caltech.edu/}).\\
c) \citet{2007A&A...473..399P,2005A&A...429L...5D}. Note that NGC~7552
is listed as a H\textsc{ii} Liner by \citet{2007A&A...473..399P}.
However, other evidence suggests that it is in fact a starburst galaxy
with a giant nuclear H\textsc{ii} region \citep{1994ApJ...433L..13F,
1998MNRAS.300..757F}. Similarly, NGC~7582 has also been classified as
Sy1 by other authors \citep[e.g.][]{2010IAUS..267..134R}.
\end{minipage}
\end{table*}

The GMRT data were compared with VLA data from the 1.425~GHz atlas of
the IRAS Bright Galaxy Sample \citep{1996ApJS..103...81C} in order to
retrieve the peak brightness and spectral index of sources. When the
source was not available in the atlas, and when the s/n allowed, the
GMRT data were instead combined with images from the SUMSS
\citep{1999AJ....117.1578B} at 843~MHz. For comparison with the
1.425~GHz atlas data, the GMRT images were convolved to a resolution of
$60$~arcsec. For SUMSS data, GMRT images were reconvolved to $45 \times
45 \csc|\delta|$~arcsec$^2$, where $\delta$ is the declination of each
observation \citep{2003MNRAS.342.1117M}. The higher frequency images
were then regridded to the same geometry as the GMRT images. The
1.425~GHz data had an r.m.s.\ noise of 0.2~mJy~beam$^{-1}$, and the
843~MHz data had an r.m.s.\ noise of $2.2$~mJy~beam$^{-1}$ in the field
of USCG~S063, and $1.8$~mJy~beam$^{-1}$ in the field of the Grus
Quartet. All compact sources in the images appear aligned to
$\le1\farcs2$.

Note that errors in the derived spectral indices are otherwise dominated
by flux calibration errors, which are assumed to be $5$\% and $3$\% for
the GMRT/VLA and SUMSS data respectively. There are further systematics
originating from the accuracy of the beam model at the GMRT. All images
were corrected for the effects of primary beam attenuation using
\textsc{pbcor} in AIPS and the coefficients that have been determined at
the observatory\footnote{\texttt{http://ncra.tifr.res.in:8081/\char'176
ngk/primarybeam/beam.html}}.

\begin{figure}
\centering
\includegraphics[clip=false, trim=0cm 0cm 0cm 0cm, width=8.4cm]{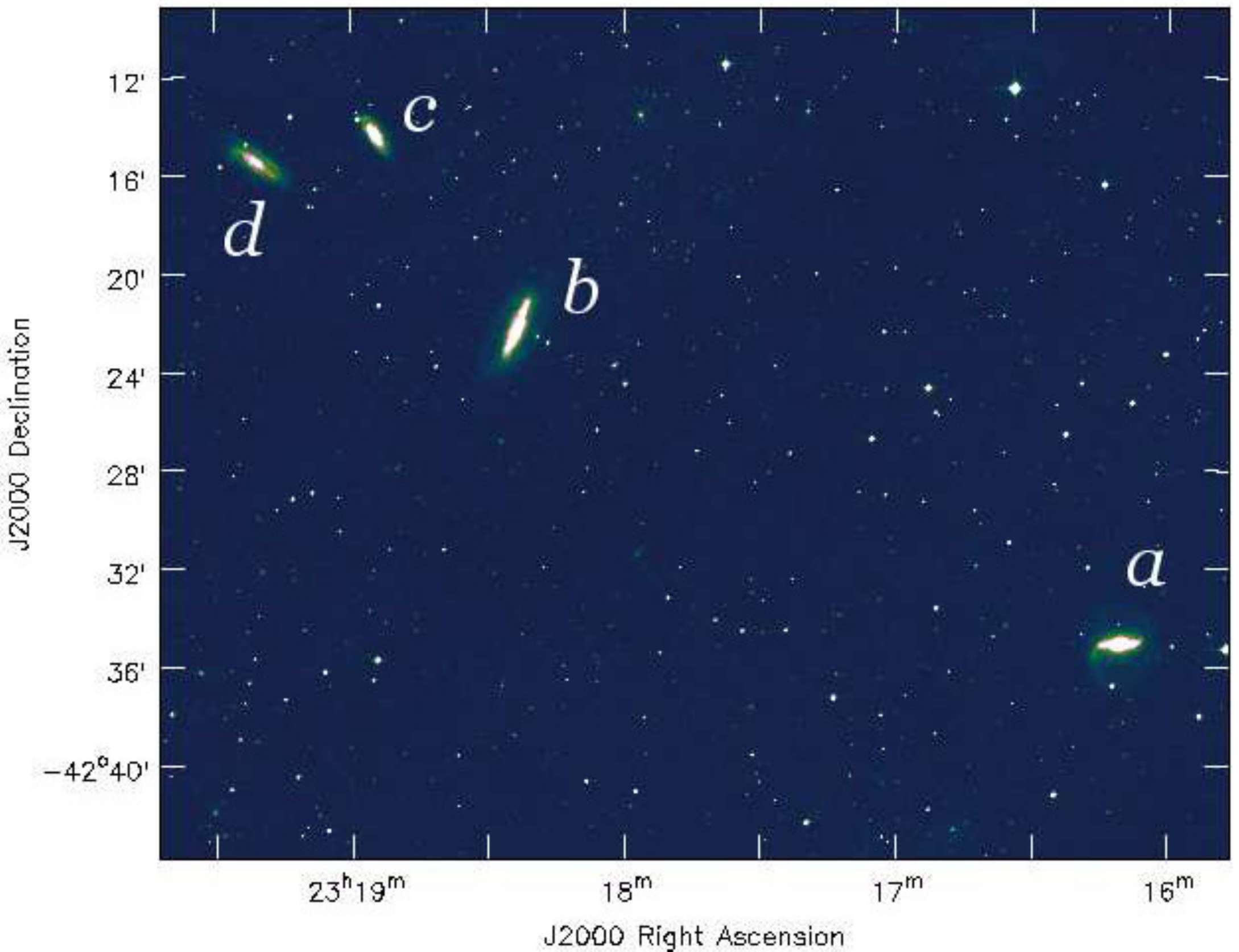}
\caption{The near-IR DSS-2 image of the four group galaxies in the Grus
Quartet. The labels are defined in Table~\ref{tab:groupinfo}. The pseudocolour scale uses the cubehelix colour scheme \citep{2011BASI...39.289G}.
}\label{SCG2-optical}
\end{figure}

\begin{figure*}
\centering
\includegraphics[clip=false,trim=0cm 0cm 0cm 0cm,width=17cm]{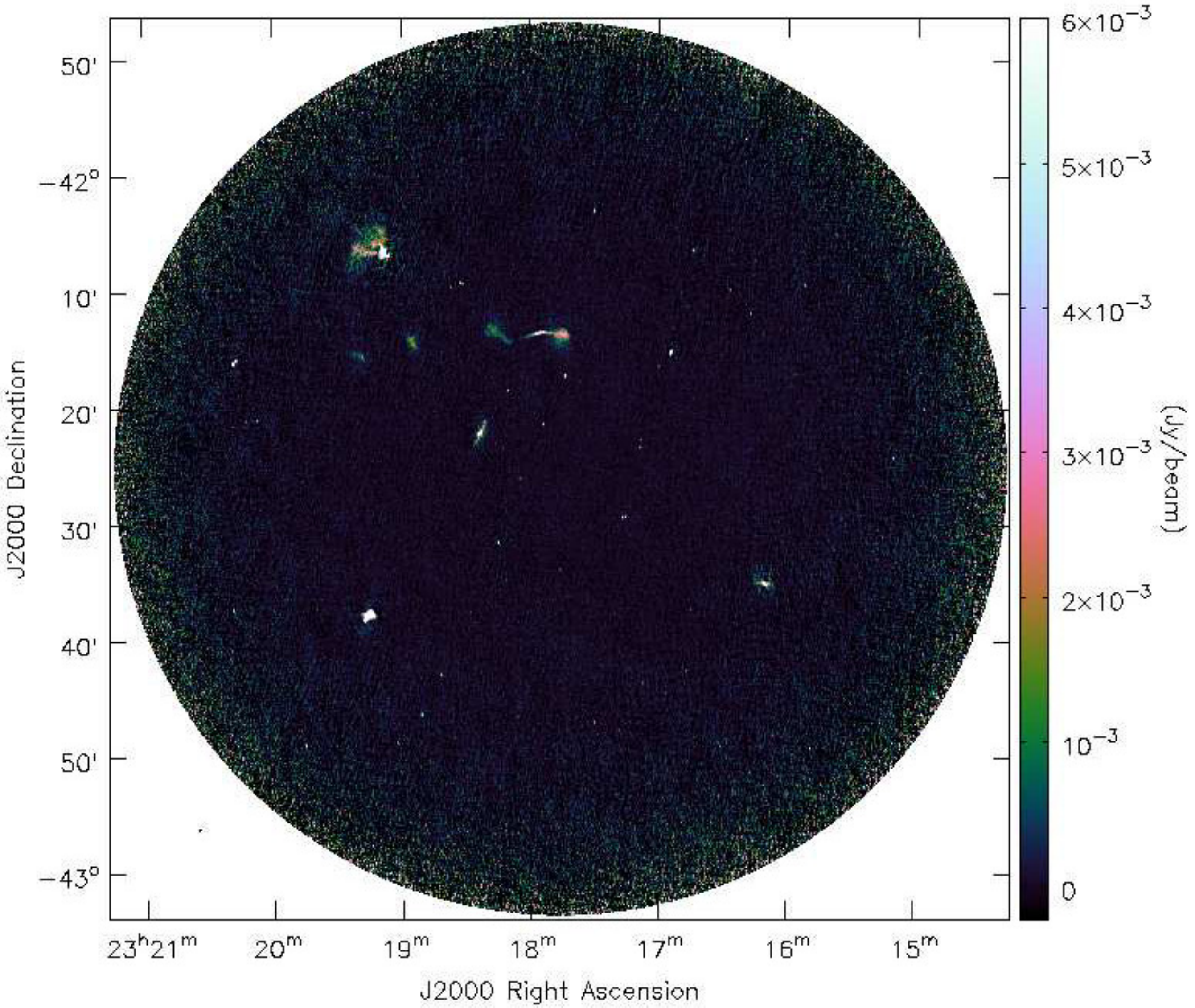}
\caption{The Stokes $I$ GMRT image of the field surrounding the Grus
quartet at full-resolution of $9\farcs5 \times 4\farcs6$ with a position
angle of $16^{\circ}$.
}\label{SCG2-PBCOR}
\end{figure*}

\section{Group Galaxies}\label{S:groupgalaxies}

The SCG catalogue consists of 121 groups over $\sim 25$\% of the
southern sky \citep{2002AJ....124.2471I}. The automated algorithm used
to select the SCG sample followed criteria that was similar to that used
for the Hickson's Compact Groups (HCGs), which consists of many groups
that are physically bound and in different stages of evolution
\citep{1982ApJ...255..382H}. The SCG catalogue avoids one of the biases
of the HCG catalogue -- as fainter galaxy members with a magnitude close
to the cut-off limit were discarded by the eyeball search used for the
identification of HCGs. The selection criteria for the SCGs are based on
richness, isolation, and compactness in order to identify groups in the
blue plates of the COSMOS survey \citep{2002AJ....124.2471I}.

The groups observed in this paper, the Grus Quartet and USCG~S063 (also
known as SCG~2315$-$4241 and SCG~0141$-$3429 respectively), are both in
`phase 2' of their evolution, i.e.\ with extended tidal tails and a
perturbed gas distribution \citep{2005A&A...429L...5D,
2007A&A...473..399P}. They are observed with the GMRT at 610~MHz and
used alongside Sydney University Molonglo Sky Survey data at 843~MHz,
VLA data at 1.425~GHz, and near infra-red data from the DSS-2 survey.
These data allow for constraints on the group members' spectral index,
polarisation fraction, and Faraday depth. They also allow for
identification of counterpart sources at other wavelengths, and act as a
probe of tidal interactions or ram pressure from the IGM compressing the
magnetic field of the galaxies.

Tidal interactions and/or ram pressure stripping have been shown to
occur frequently in compact groups, with $\sim 50$\% of the galaxies
showing signs of morphological disturbances. In a simple case, tidal
interactions would lead to bridges and tails, ram pressure stripping
would lead to a smooth swept-up morphology and displaced gas disks,
while turbulent viscous stripping would lead to effects on the entire
gas disk \citep{1994AJ....107.1003C}. It is likely that these
hydrodynamical interactions play a substantial role in the evolution of
group galaxies and their morphological type
\citep[e.g.][]{2007A&A...473..399P}. Several groups of galaxies have
shown evidence for ram pressure like events shaping their morphology,
with many member galaxies showing an {\sc H\,i} deficiency, alongside
the presence of massive {\sc H\,i} structures in the IGM
\citep[e.g.][]{2001A&A...377..812V, 2005A&A...435..483K,
2006MNRAS.373..653R, 2006MNRAS.371..739K, 2007MNRAS.378..137S,
2009MNRAS.400.1962K, 2012ApJ...747...31R, 2013arXiv1302.0285M}. The
presence and role played by the IGM in the evolution of groups of
galaxies therefore continues to be of interest
\citep[e.g.][]{1996AAS...189.8305M}. X-ray emission has been detected
from the IGM of several groups \citep[e.g.][]{2003ApJS..145...39M}, with
the detection rate being higher in the evolved elliptical dominated
groups as compared to spiral-rich groups. However, as shown by
\citet{2005A&A...435..483K}, the member galaxies of the poor group
Holmberg 124 seem to show signatures attributable to ram pressure like
events being experienced by the galactic interstellar medium.
Furthermore, cold {\sc H\,i} gas has been detected between the member
galaxies in several Hickson's groups \citep{2001A&A...377..812V}, which
could enhance the density of the IGM and lead to ram pressure like
events playing a more significant role. Nevertheless, due to
observational difficulties and faint signatures, it has been difficult
to rigorously determine the effect of the IGM on the galaxies in groups.

Another way to address this is to study the linearly polarised emission
from the group members. Polarised emission is, in principle, an ideal
tracer of such interactions, and directly reveals regular magnetic
fields that result from field compression due to interaction with the
IGM \citep{2004A&A...421..571H}. Previous attempts to understand these
processes have been made both observationally
\citep[e.g.][]{2012A&A...545A..69W, 2013arXiv1304.1279V} and through
three-dimensional magnetohydrodynamical simulations
\citep{2010NatPh...6..520P}. Such a study has been done for spiral
members of the Virgo cluster \citep{2007A&A...464L..37V,
2013arXiv1304.1279V}, with all the studied galaxies showing an
asymmetric polarised intensity distribution along ridges in the outer
part of each galaxy which differs from what is typically observed in
field spiral galaxies. \citet{2007A&A...464L..37V} attributed the
observed behaviour of these galaxies to interactions with the
intracluster medium that are mediated by ram pressure. While
\citet{2013arXiv1304.1279V} found that tidal interactions and accreting
gas envelopes can lead to compression and shear motions which enhance
the polarised radio continuum emission. Observing a large sample of
galaxy groups in linear polarisation will therefore assist in further
understanding the signatures of various processes that influence the
radio emission.

Various properties of the galaxies in the target groups are summarised
in Table \ref{tab:groupinfo}. The individual groups are discussed in
further detail in Sections \ref{group2} and \ref{group1}.

\subsection{The Grus Quartet}\label{group2}

The Grus Quartet at a luminosity distance of 18.3~Mpc and a velocity of
1571~km~s$^{-1}$ contains four galaxies, all of which are spirals as
shown in Fig.~\ref{SCG2-optical}. X-ray emission has previously been
detected from three of the galaxies -- NGC~7552, NGC~7582, and NGC~7590
\citep{1978ApJ...223..788W,1981ApJ...246L..11M}.

The four members of the Quartet have been observed in {\sc H\,i} by
\citet{2005A&A...429L...5D} who found an extended tidal tail with an
{\sc H\,i} mass of $1.34\times10^9$~$\textrm{M}_{\sun}$ emerging from
NGC~7582 and also a large {\sc H\,i} cloud of mass
$7.7\times10^8$~$\textrm{M}_{\sun}$ removed from the galaxy and now a
part of the IGM. The group is evolving with 11\% of its {\sc H\,i} in
the IGM \citep{2005A&A...429L...5D}, making it an important case study
of how the enhanced densities in the IGM might be affecting the other
properties of the galaxy.

\begin{figure}
\centering
\hspace{-28pt}\includegraphics[clip=false,trim=0cm 0cm 0cm 0cm,height=7.0cm]{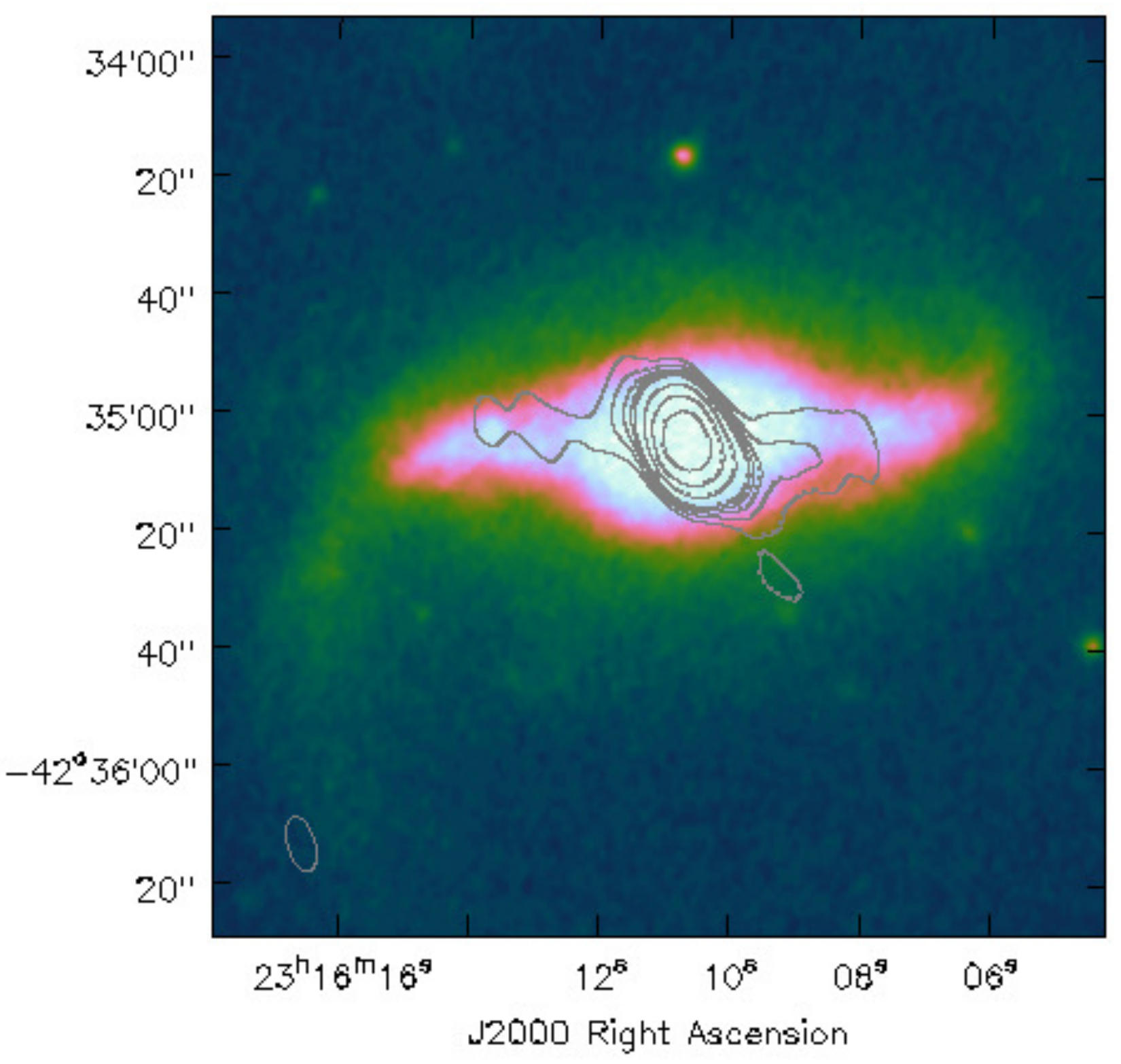}
\bigskip
\hspace{0pt}\includegraphics[clip=false,trim=0cm 0cm 0cm 0cm,height=7.0cm]{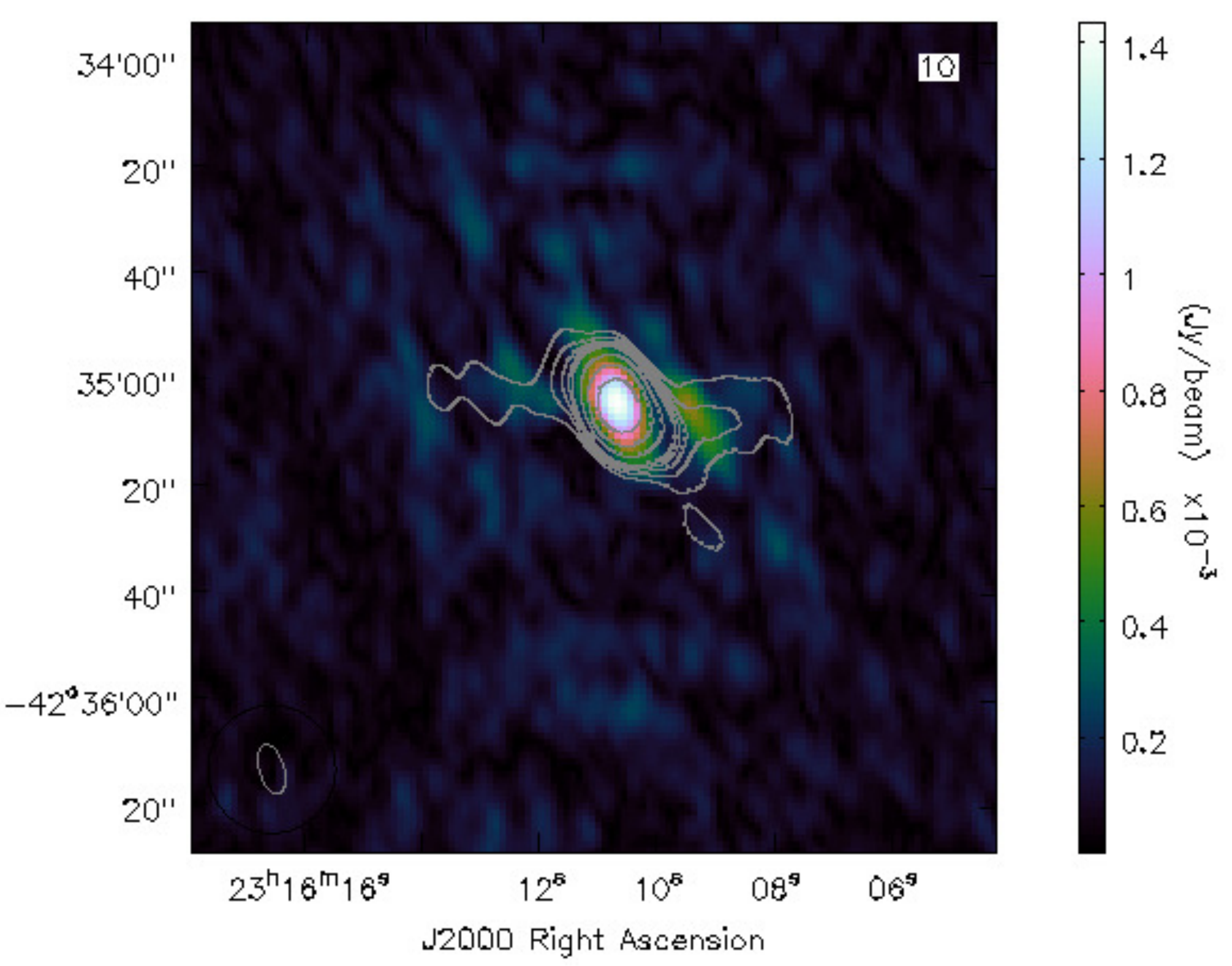}
\caption{Top: Stokes $I$ contours at 610~MHz overlaid on the near-IR
DSS-2 image of NGC~7552, in the field of the Grus Quartet.
The contours are at ($-2$, $-1$, 1, 2, 3, 4, 5, 10, 20, 40, 80) $\times$
the off-source $4\sigma$ level, where $\sigma$ is
295~$\muup$Jy~beam$^{-1}$. The 610~MHz image is at full resolution of
$9\farcs5 \times 4\farcs6$ with a PA of $16^{\circ}$. The synthesised beam is shown in the bottom left. Bottom: Stokes $I$
contours at 610~MHz overlaid on the $P$ image of
NGC~7552 at a Faraday depth of $10$~rad~m$^{-2}$. The $P$ image has a resolution of $24$~arcsec, and has not
been corrected for the effects of the primary beam or for Rician
bias.
}\label{SW-SCG2-galaxy}
\end{figure}
\begin{figure}
\centering
\hspace{-38pt}\includegraphics[clip=true, trim=1.0cm 0.9cm 0cm 0cm, height=8.5cm]{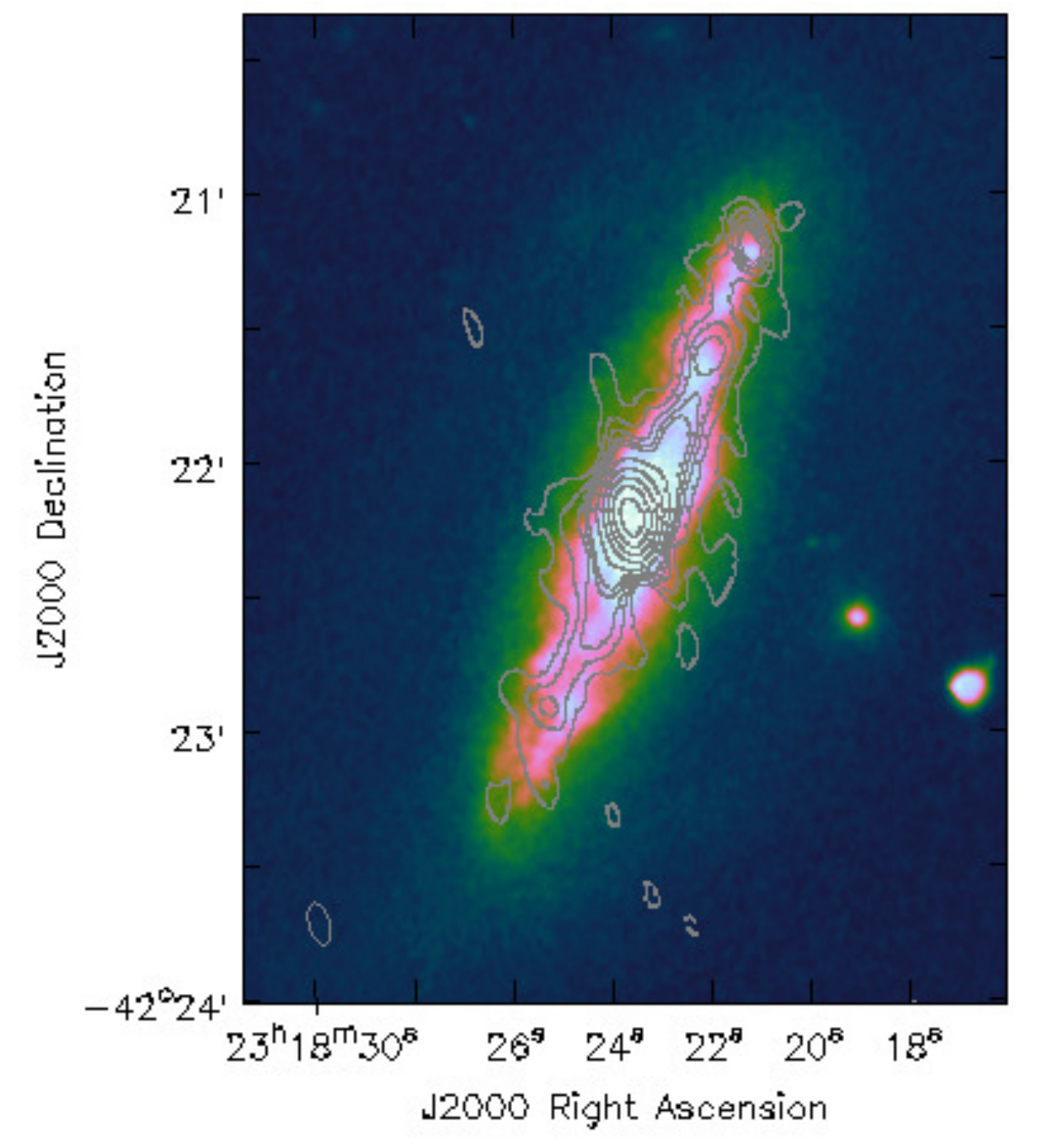}
\hspace{0pt}\includegraphics[clip=true, trim=1.2cm 0.9cm 0cm 0cm, height=8.5cm]{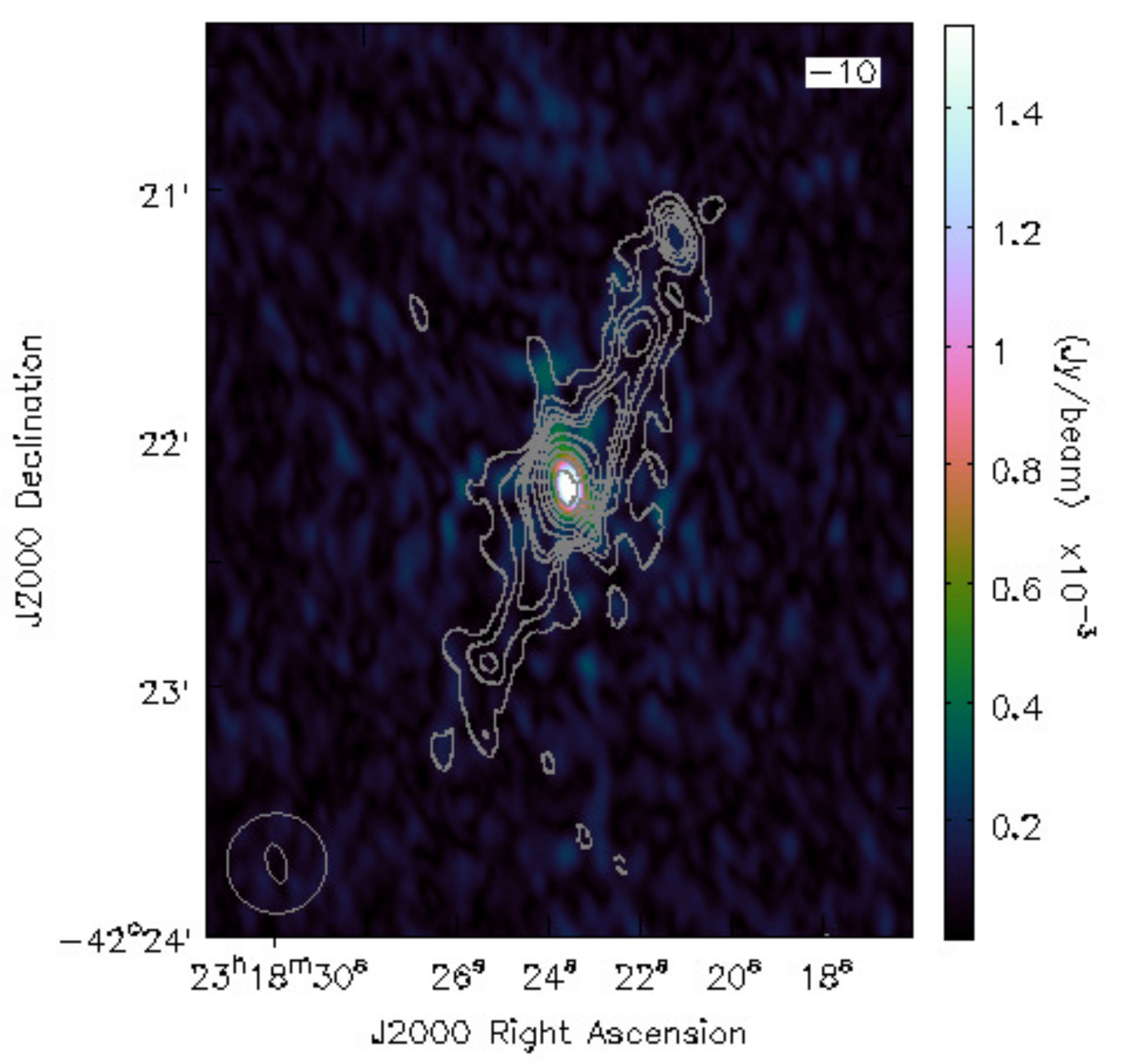}
\caption{Top: Stokes $I$ contours at 610~MHz overlaid on the near-IR
DSS-2 image of NGC~7582 in the field of the Grus Quartet. The contours
are at ($-2$, $-1$, 1, 2, 3, 4, 5, 10, 20, 40, 80, 160) $\times$ the
off-source $4\sigma$ level, where $\sigma$ is 185~$\muup$Jy~beam$^{-1}$.
The 610~MHz image is at full resolution of $9\farcs5 \times 4\farcs6$
with a PA of $16^{\circ}$. Bottom: Stokes $I$ contours at 610~MHz
overlaid on the $P$ image of NGC~7582 at a Faraday depth of $-10$~rad~m$^{-2}$. The $P$
image has a resolution of $24$~arcsec, and has not been corrected for
the effects of the primary beam or for Rician bias. The synthesised beams are shown in the bottom left.}\label{Centre-SCG2-galaxy}
\end{figure}

The 610~MHz data from observation 17\_060\_2 achieved $\sigma =
85~\muup$Jy~beam$^{-1}$ near the phase-centre in Stokes $I$ at
full-resolution of $9\farcs5 \times 4\farcs6$ with a position angle of
$16^{\circ}$, and a noise level in the band-averaged $Q/U$ of $\sim
36~\muup$Jy~beam$^{-1}$ at a resolution of $24$~arcsec. The low
resolution for the polarisation images were chosen in order to maximise
the s/n within each beam. The removal of antennas with leakages $>15$\%,
some of which were located in the GMRT's outer arms, meant that the data
had to be tapered in the $uv$-plane in order to avoid large gaps in the
$uv$-coverage. The cleaned $\phi$-cubes had
$\sigma=54~\muup$Jy~beam$^{-1}$~rmsf$^{-1}$. The Stokes $I$ FOV is shown
in Fig.~\ref{SCG2-PBCOR}.

\begin{figure}
\centering
\includegraphics[clip=true, trim=0.9cm 0.9cm 0cm 0cm, width=8.5cm]{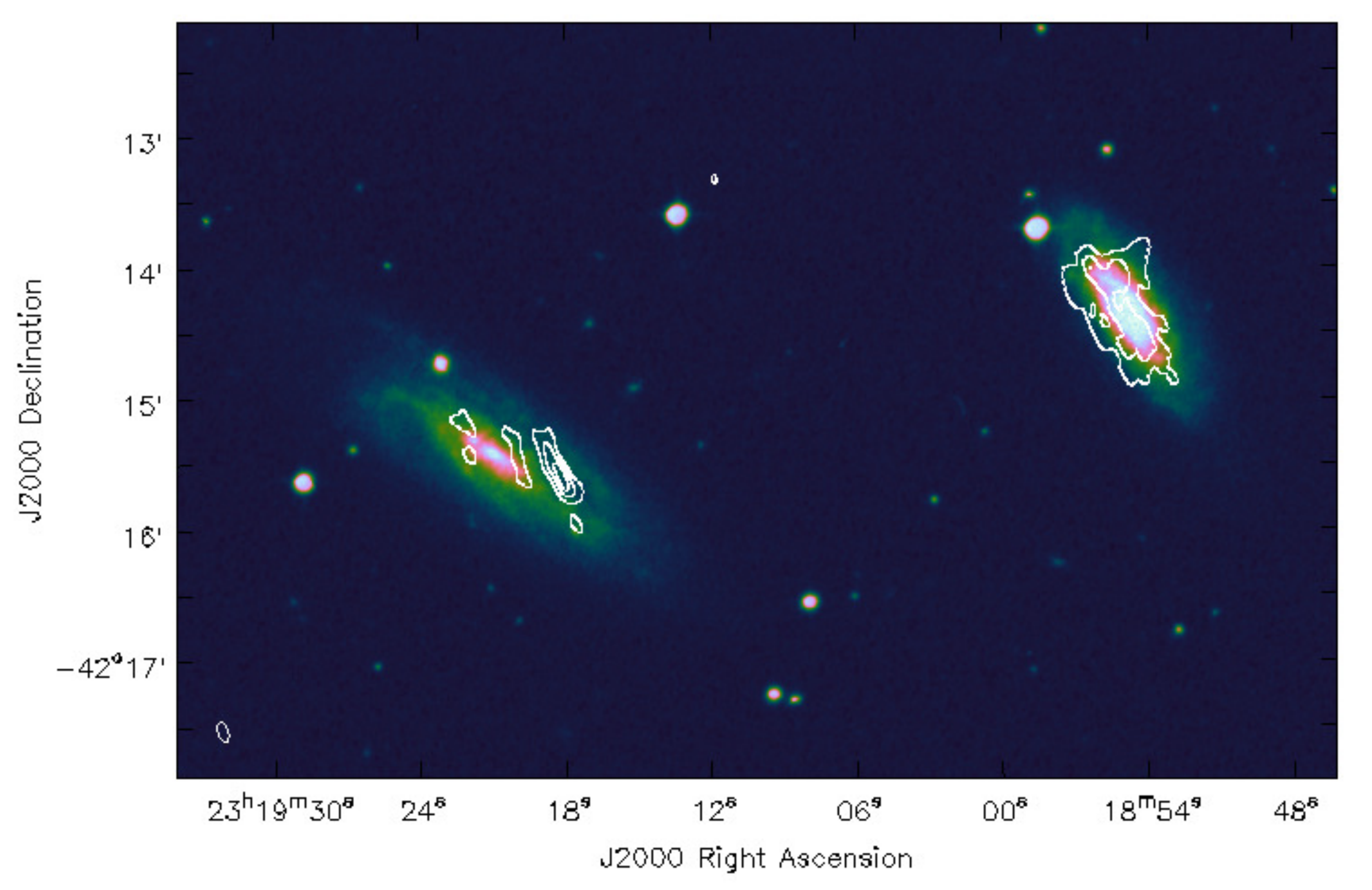}
\caption{Stokes $I$ contours at 610~MHz overlaid on the near-IR
DSS-2 image of NGC~7590 and NGC~7599 in the field of the Grus Quartet.
NGC~7590 is further to the west. The contours are at ($-2$, $-1$, 1, 2,
3) $\times$ the off-source $4\sigma$ level, where $\sigma$ is
205~$\muup$Jy~beam$^{-1}$. The 610~MHz image is at full resolution of
$9\farcs5 \times 4\farcs6$ with a PA of
$16^{\circ}$. The synthesised beam is shown in the bottom left.}\label{NE-SCG2-galaxies}
\end{figure}

At 610~MHz, all four group galaxies were detected in Stokes $I$. The galaxies NGC~7552, NGC~7582, NGC~7590, and NGC~7599 were found to have a peak brightness of 377, 379, 75, and 53~mJy~beam$^{-1}$ respectively. The Stokes $I$, near-IR data, and polarised intensity at 610~MHz are shown for NGC~7552 in Fig.~\ref{SW-SCG2-galaxy}, and for NGC~7582 in Fig.~\ref{Centre-SCG2-galaxy}. The emission from NGC~7590 and NGC~7599 is shown in Fig.~\ref{NE-SCG2-galaxies}.

The brightest emission emanates from the nuclei of NGC~7552 and
NGC~7582. Extended emission is detected at 610~MHz along the bar of
NGC~7552. There is clear radio structure along the bar of NGC~7582
through to the outer northern and southern edges of the galaxy. These
knots of emission are also visible in H$\alpha$
\citep{1999AJ...118..730H}. Diffuse emission is detected across most of
the disk of NGC~7590, with very little structure in the emission. The
emission from NGC~7599 is faint, and only the peaks in emission are
detected. The peak brightness in NGC~7599 is located towards the west of
the outer disk, and arises from a region associated with the spiral arm
which emerges in the south and then turns towards the north.

There was insufficient resolution to investigate spatial spectral index
variations across these galaxies. For the brightest emission,
$\alpha_{610}^{1425}$ was found to be $0.62\pm0.05$ in NGC~7552,
$0.62\pm0.05$ in NGC~7582, $0.71\pm0.24$ in NGC~7590, and $0.97\pm0.35$
in NGC~7599. This emission comes from the cores of NGC~7552 and
NGC~7582, and is integrated across the disk in both NGC~7590 and
NGC~7599. The knot of emission in the bar to the north of NGC~7582 (and
correlated with H$\alpha$) is just resolved, and has $\alpha=0.8\pm0.3$.

\begin{figure}
\centering
\includegraphics[clip=true,trim=2cm 0cm 0.2cm 0cm,width=8.4cm]{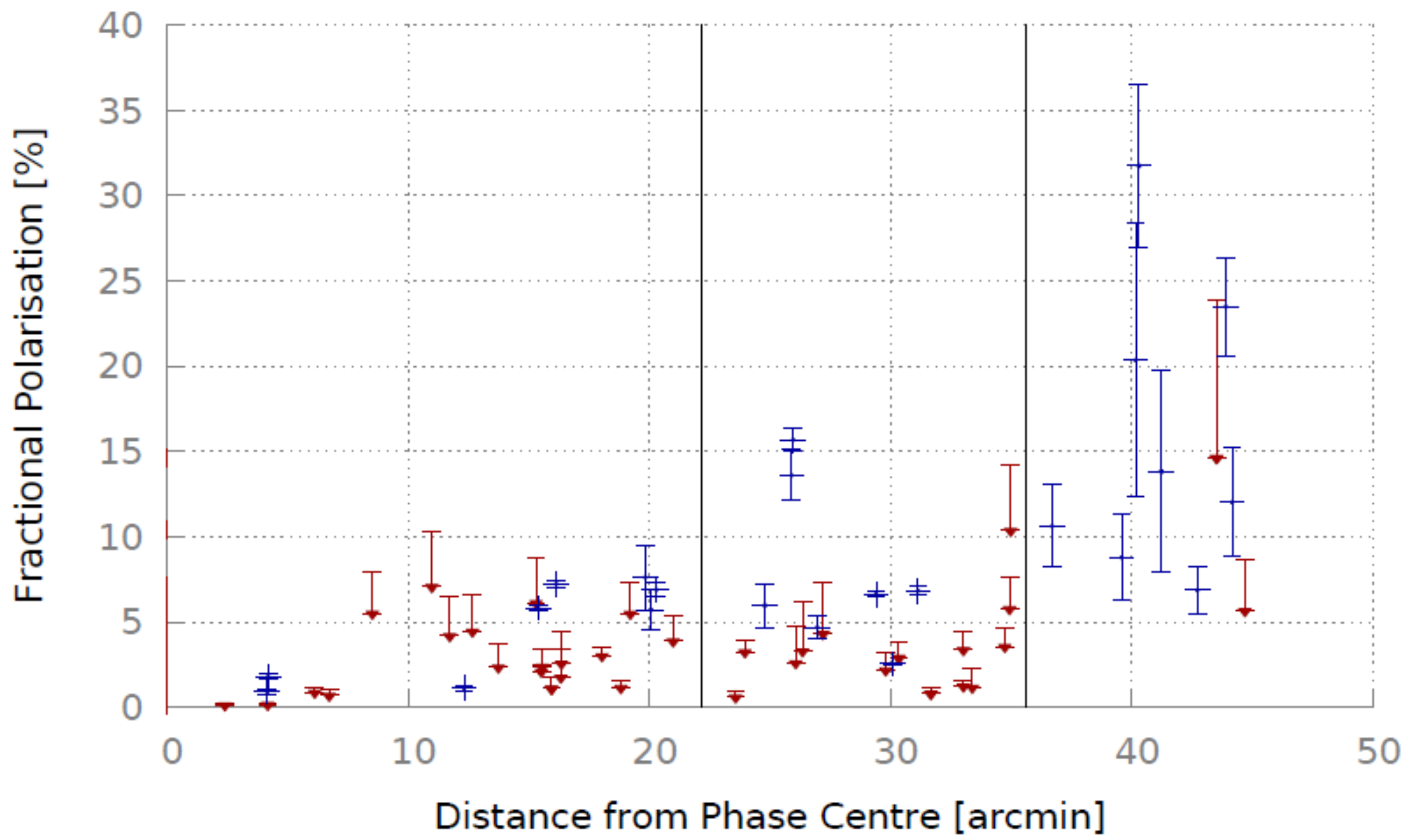}
\caption{The fractional polarisation, $\Pi$, of sources surrounding the
nearby galaxy M51 as a function of distance from the phase-centre for a
full-track observation with considerable averaging down over the large
range in parallactic angle. Sources detected above an $8\sigma$ limit
following RM Synthesis are shown with error bars. Sources undetected in
polarisation within the sensitivity limit ($P < 8\sigma$) are shown as
an upper limit. The half-power and $10\%$ points of the $I$ beam are
indicated by the solid vertical lines
\citep{FarnesRMAA}.
}\label{offaxis}
\end{figure}

NGC~7590 and NGC~7599 were undetected in polarisation to $3\sigma$ upper
limits of $<0.9$\% and $<1.4$\% respectively. Bright polarisation was
detected from the nucleus of both NGC~7552 and NGC~7582. It is important
to compare this polarised emission to that expected due to
direction-dependent instrumental effects -- especially for these two
`polarised' galaxies which are both displaced from the phase-centre.
This comparison is complicated by a number of factors, especially since
the amount of spurious polarisation is reliant on both the position of
the source within the primary beam and the precise values of source and
instrumental polarisation at this location. It is important to note that
for observations with a long track in parallactic angle, the
direction-dependent instrumental polarisation averages down
substantially, as shown in Fig.~\ref{offaxis}. This averaging is limited
for these observations due to the low declination. USCG~S063 was
observed with parallactic angles ranging from $-32^{\circ}$ to
$+51^{\circ}$, while the Grus Quartet was observed with parallactic
angles ranging from $-29^{\circ}$ to $+31^{\circ}$. Some averaging down
of the off-axis response is therefore expected, particularly as
measurements of the polarisation beam indicate that it is oriented
radially outwards.

The polarised peak brightness and Faraday depth were measured by fitting
a Gaussian to the points surrounding the peak of the Faraday spectrum.
Only pixels with a peak in $\phi$-space greater than $8\sigma$ were
analysed, where $\sigma$ is the noise in the cleaned $\phi$-cubes. The
polarised intensity measurements were corrected for the effects of
Rician bias using an alternative estimator, $P_{0} = \sqrt{P^2 -
2.3\sigma_{QU}^2}$, which is a more effective estimator of the true
polarised intensity in $\phi$-cubes \citep{2011arXiv1106.5362G}, where $P$ is the linearly polarised intensity, and $P_{0}$ is the corrected quantity.

The corrected values were then used to calculate the fractional
polarisation, $\Pi = P_{0}/I$. The fractional polarisations are
$0.87\pm0.02$\% and $0.70\pm0.01$\% for NGC~7552 and NGC~7582
respectively. Note that the quoted 1$\sigma$ uncertainties are our measurement errors -- there are likely further contributions due to systematics in the calibration of the instrument. We estimate the maximum residual instrumental polarisation at the phase-centre to be $\le0.25$\%. The Stokes $I$, near-IR data, and polarised intensity at 610~MHz are shown for NGC~7552 in Fig.~\ref{SW-SCG2-galaxy}, and for NGC~7582 in Fig.~\ref{Centre-SCG2-galaxy}.

\begin{figure}
\centering
\includegraphics[clip=true, trim=1.1cm 0.2cm 0.1cm 0cm,
   width=8.4cm]{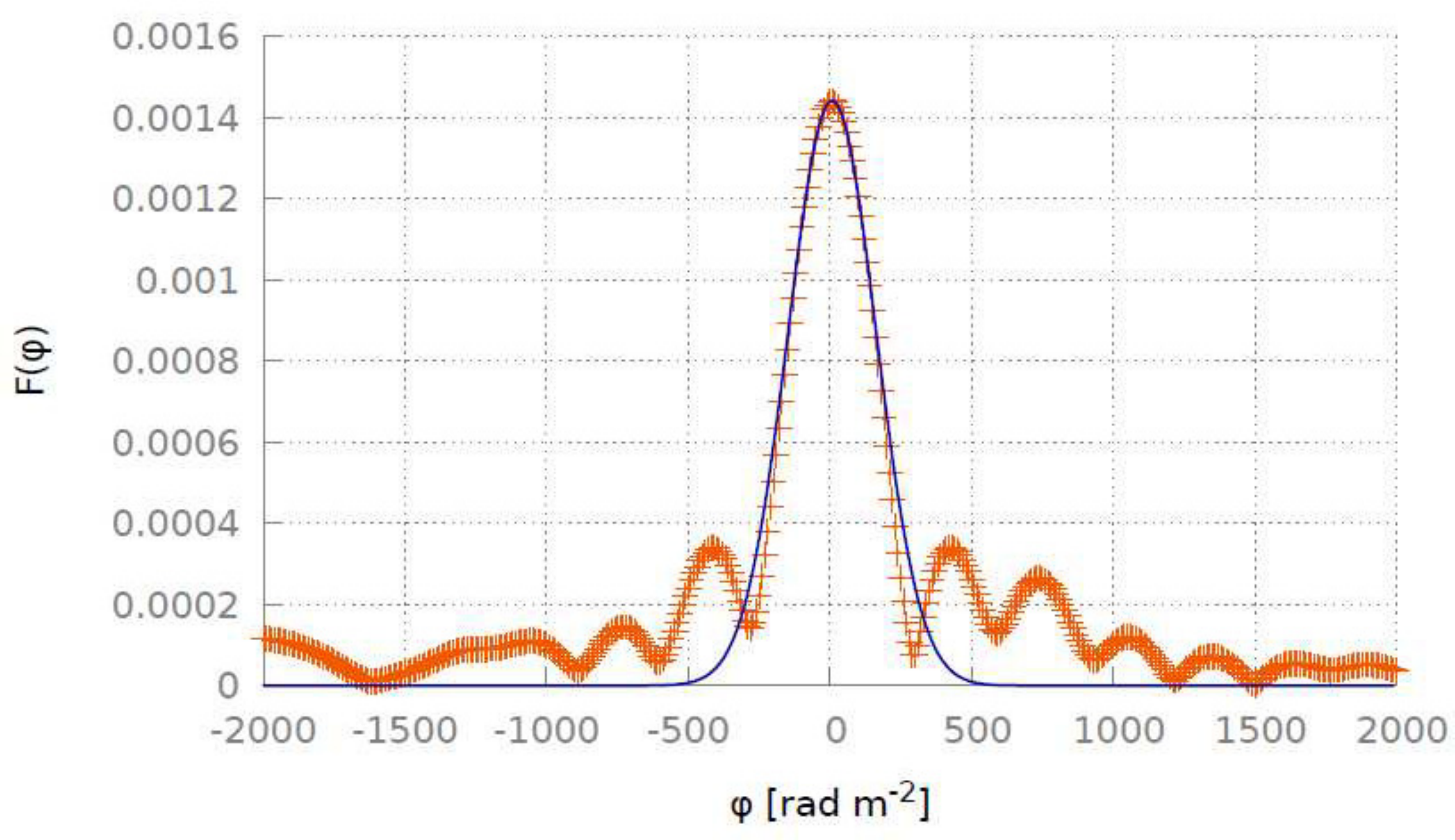}
\bigskip
\includegraphics[clip=true, trim=1.1cm 0.2cm 0.1cm 0cm,
   width=8.4cm]{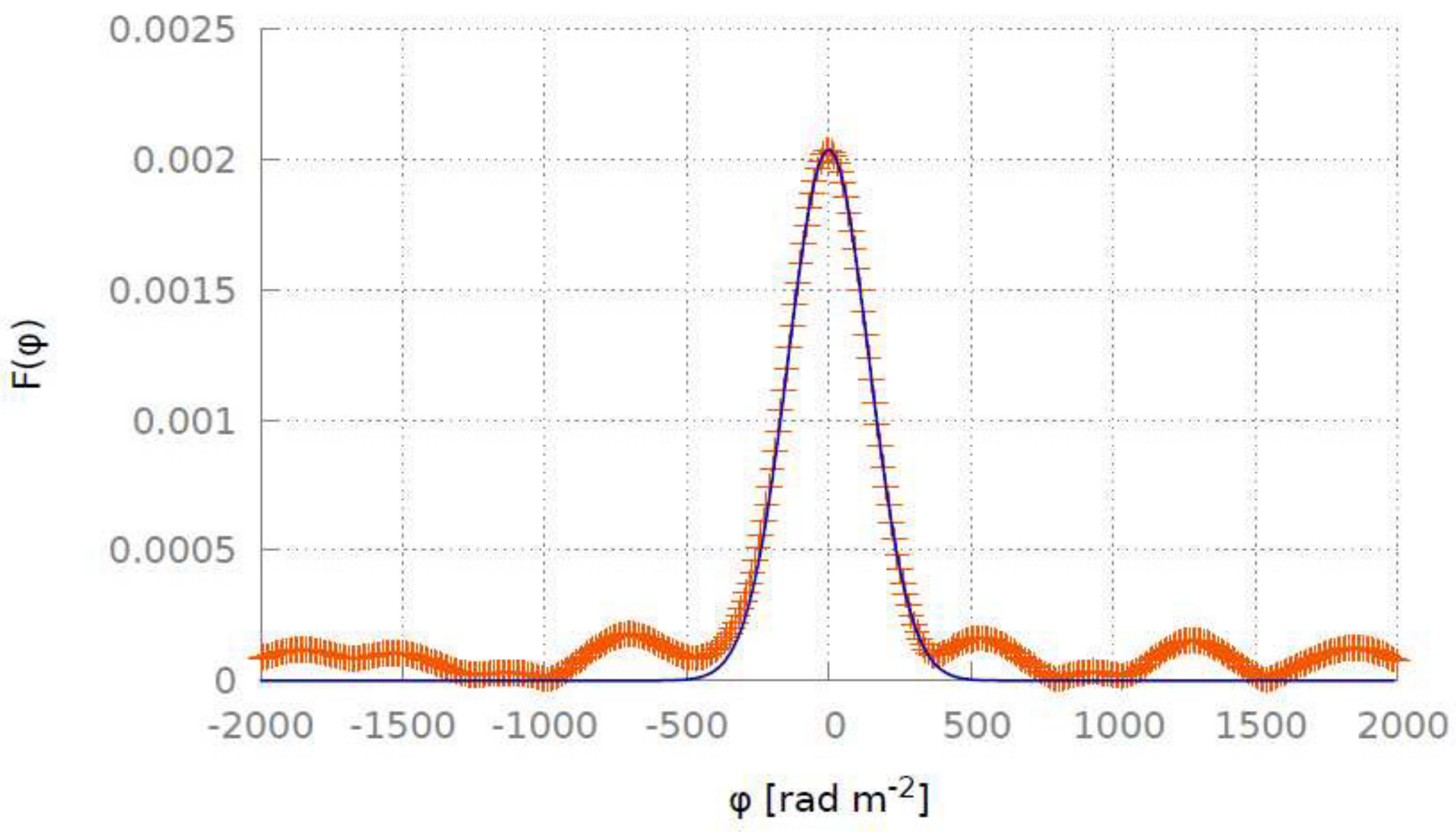}
\caption{A plot of the Faraday dispersion function as a function of
Faraday depth for the brightest polarised pixel in both NGC~7552 (top)
and NGC~7582 (bottom) at $24$~arcsec resolution. The datum extracted at
each trial Faraday depth is shown in light grey. A Gaussian fit to the
peak of the spectrum is shown by the solid line. Note that the vertical
scales are different.
}\label{GalaxyA-Faraday_spectrum}
\end{figure}

NGC~7552 and NGC~7582 are located $20\farcm1$ and $7\farcm6$ from the
phase-centre, yielding an approximate upper limit for the direction-dependent
instrumental polarisation of $2.0$\% (NGC~7552) and $0.3$\% (NGC~7582).
The estimated upper limit at the location of NGC~7552 is large enough to
account for the polarisation detection as an instrumental effect,
although the effects of parallactic angle averaging cannot be trivially
taken into account. As the frequency-dependence of the instrumental polarisation has been removed from the data (see Section~\ref{onaxisleakagesection}) and the polarisation beam has been found to be essentially independent of frequency across the GMRT band (see Section~\ref{offaxispol}), instrumental effects will tend to accumulate at
Faraday depths (FDs) of $0$~rad~m$^{-2}$ \citep[see e.g.][for further details]{2005A&A...441.1217B}. RM Synthesis was therefore applied in order to diagnose the nature of the emission.
The peak in the Faraday Dispersion Function is at a FD of
$7.71\pm0.29$~rad~m$^{-2}$ -- a plot of the Faraday spectrum is shown in
Fig.~\ref{GalaxyA-Faraday_spectrum}. Surrounding pixels have varying FDs
from $\sim4{-}14$~rad~m$^{-2}$. The Faraday depth (FD) of the pixel of
peak polarisation was found to be $7.8\pm0.5$~rad~m$^{-2}$. The
polarisation to the west of the nucleus has a `smudged' appearance and
does not coincide with a distinct feature of the galaxy in Stokes $I$.
It is likely that errors in the determination of the leakage phase
arising from the use of a linear model have led to this spurious
polarisation to the west of the nucleus. As the FD is non-zero, it implies that the polarisation detection does not result from the effects of instrumental polarisation.

Following RM Synthesis of the brightest polarised pixel in NGC~7582, the
peak in Faraday space occurs at a FD of $-1.83\pm0.22$~rad~m$^{-2}$ -- a
plot of the Faraday spectrum is shown in
Fig.~\ref{GalaxyA-Faraday_spectrum}. FDs in the immediately surrounding
bright pixels are all negative, varying between $-9$~rad~m$^{-2}$ to
$-1$~rad~m$^{-2}$. The consistently negative, and non-zero, FDs are again indicative of the detection not resulting from instrumental polarisation.

\begin{figure}
\centering
\includegraphics[clip=true, trim=0.9cm 1.2cm 0.3cm 0.0cm, width=8.5cm]{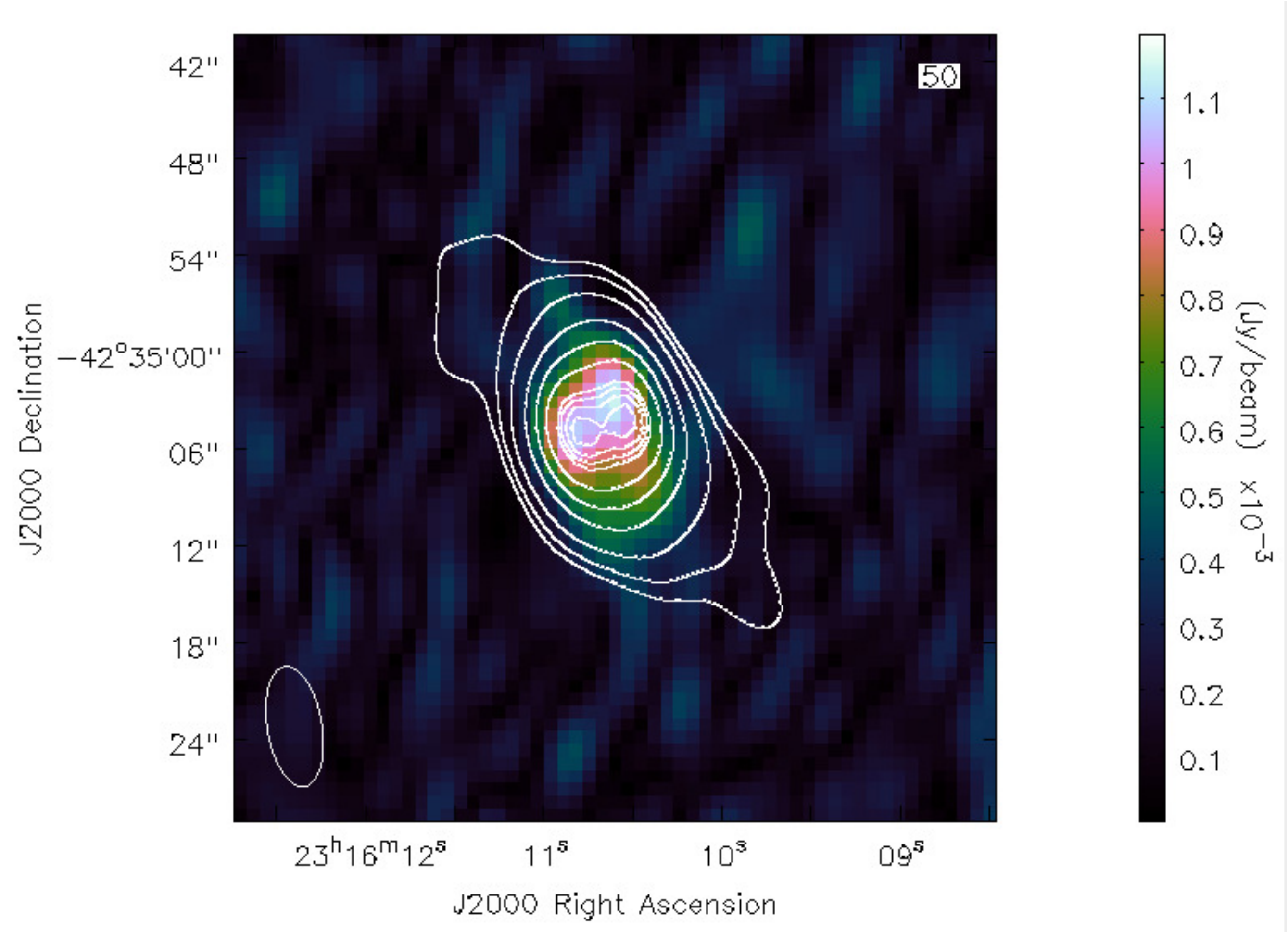}
\caption{The high-resolution image of the core of NGC~7552. Stokes $I$
contours at 610~MHz are overlaid on the RM Synthesis image of $P$ at a
Faraday depth of $50$~rad~m$^{-2}$. The image has a resolution of
$7\farcs5 \times 3\farcs4$ and a position angle of $9\fdg5$. The
contours are at (1, 2, 4, 8, 12, 16, 20, 21, 22, 23) $\times$ the
off-source $3\sigma$ level. Both the total intensity and polarisation
images are uniformly weighted. The synthesised beam is shown in the bottom left.}\label{starburstring}
\end{figure}

In an attempt to constrain the presence of beam depolarisation, the data
were over-resolved using uniform weighting. This provided a beam with a
FWHM of $7\farcs5 \times 3\farcs4$ and a position angle of $9\fdg5$. A
cellsize of $0\farcs8$ was used to ensure the synthesised beam was
well-sampled. The core of NGC~7582 remained unresolved in both total
intensity and linear polarisation. However, the core of NGC~7552 reveals
a structure that is just-resolved on these scales (with $\approx2$~beams
across the source). The polarisation structure, overlaid with total intensity contours, is shown in Fig.~\ref{starburstring}. Note that no polarisation was detected in the arms
of the galaxy, as has been previously observed at higher frequencies
\citep{2002ASPC..275..331B,2002ASPC..275..361E}. The presence of a
circumnuclear starburst ring in this galaxy has been previously
well-documented \citep{1994ApJ...433L..13F,1998MNRAS.300..757F}.
Circumnuclear starburst rings can exhibit relatively large magnetic
field strengths in comparison to the rest of a galaxy -- for example,
the ring in NGC~1097 has a field strength of $60$~$\muup$G
\citep{2005A&A...444..739B}. While no AGN has been confirmed in
NGC~7552, if there was one then material could be transported from the
ring to the central black hole by magnetic braking as believed to be the
case in NGC~1097 \citep{1999Natur.397..324B}.

The polarisation image reveals two hotspots, one located to the east (at
$23^{\rm h} 16^{\rm m} 10\fs9$, $-42^\circ 35' 4\farcs7$) and another to
the northwest (at $23^{\rm h} 16^{\rm m} 10\fs6$, $-42^\circ 35'
3\farcs1$) both of which are coincident with the regions `A' and `D'
respectively as found and defined by \citet{1994ApJ...433L..13F}. The
locations of peak polarised intensity are offset to the peaks in total
intensity. The fractional polarisation increases at high-resolution
relative to the $24$~arcsec resolution images -- consistent with the
polarisation structure changing on scales less than the original beam
size. The uniform-weighted data provide a lower limit on the fractional
polarisation of $2.60\pm0.24$\% in the eastern edge of the ring, and
$2.09\pm0.23$\% to the northwest. The percentage polarisation is similar
to that reported at higher frequencies
\citep[e.g.][]{1999Natur.397..324B}. The peak FD in these regions were
found to be $57.7\pm0.7$~rad~m$^{-2}$ to the east, and
$-3.3\pm0.6$~rad~m$^{-2}$ to the northwest -- demonstrating a change in the heading (i.e.\ a change of direction along the line of sight) of the magnetic field on either side of the ring. Follow-up VLBI observations would reveal further
information on the magnetic field structure.

\subsection{USCG~S063}\label{group1}

\begin{figure}
\centering
\includegraphics[clip=false, trim=0cm 0cm 0cm 0cm, width=8.4cm]{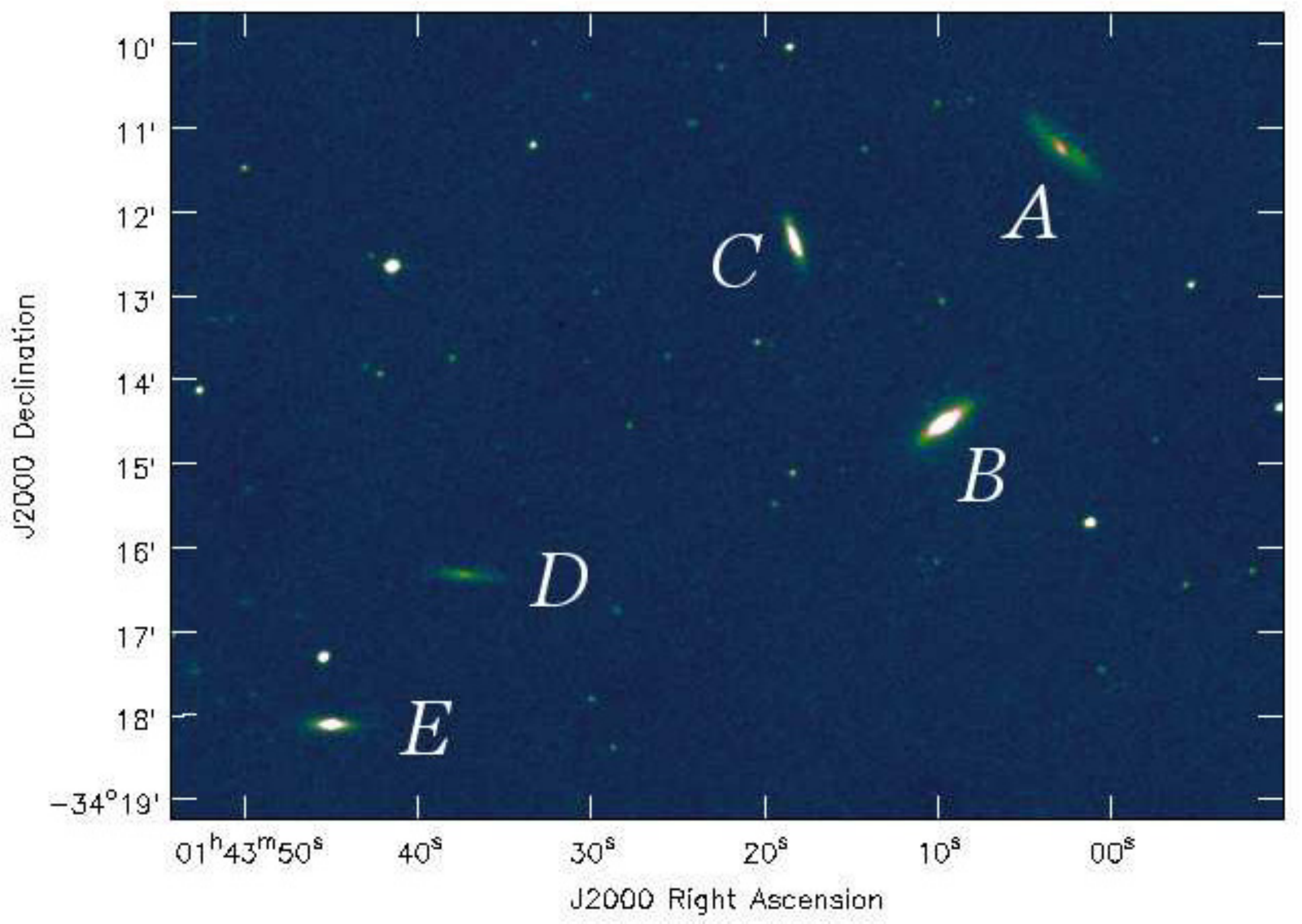}
\caption{The near-IR DSS-2 image of the five group galaxies in
USCG~S063. The labels are defined in Table~\ref{tab:groupinfo}.
}\label{SCG1-optical}
\end{figure}

The group USCG~S063 contains five galaxies, all of which are either
lenticular or spiral as shown in Fig.~\ref{SCG1-optical}.

\begin{figure*}
\centering
\includegraphics[clip=false,trim=0cm 0cm 0cm 0cm, width=15cm]{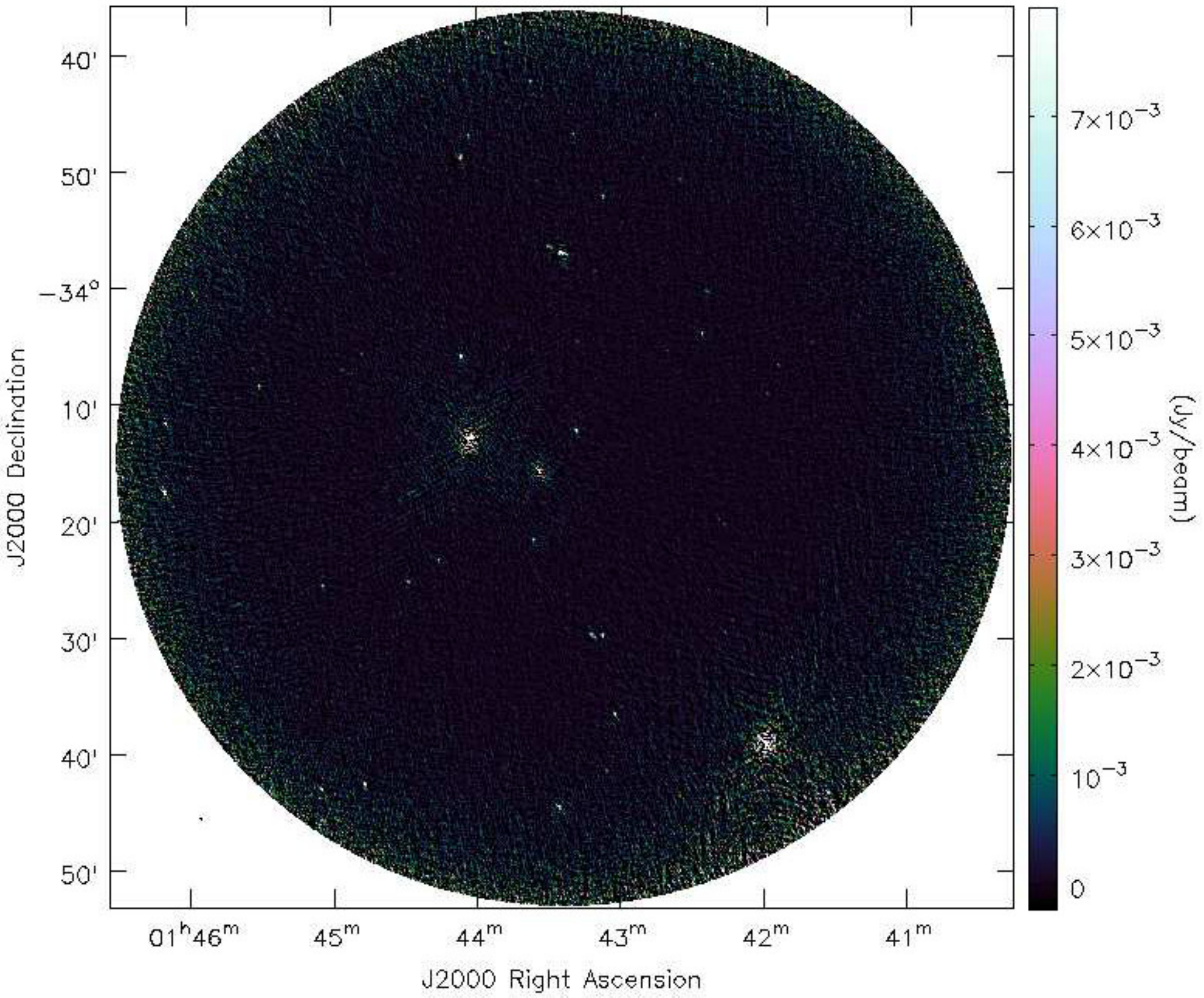}
\caption{The Stokes $I$ GMRT image of the field surrounding USCG~S063 at
full-resolution of $9\farcs7 \times 5\farcs6$ with a position angle of
$37^{\circ}$.
}\label{SCG1-PBCOR}
\end{figure*}

The 610~MHz data from observation 17\_060\_1 achieved a noise-level of
$\sigma = 270$~$\muup$Jy~beam$^{-1}$ near the phase-centre in Stokes $I$
at full-resolution of $9\farcs7 \times 5\farcs6$ with a position angle
of $37^{\circ}$. The field was strongly affected by residual phase
errors from an off-axis source located $8\farcm5$ from the phase-centre
with a peak brightness of 213~mJy~beam$^{-1}$ at full-resolution. The
algorithm \textsc{peelr} can help to remove such phase errors. Standard
self-calibration only provides a single gain solution per antenna, but
this gain is only accurate in the direction towards the calibration
source from which the solution was derived. Nevertheless, the signal
from sources at different positions in the FOV pass through different
regions of the ionosphere. This is an increasingly significant problem
at lower observing frequencies, as both the effects of the ionosphere
and the size of the FOV increase. It is therefore necessary to find
solutions for the direction-dependent gains, which can be done through
iterative `peeling' of the phase solutions at each position. Application
of \textsc{peelr} led to a mild improvement in the residual phase
errors, but the off-axis source limits the obtainable dynamic range in
the surrounding region. The cleaned $\phi$-cubes had
$\sigma=43$~$\muup$Jy~beam$^{-1}$~rmsf$^{-1}$ at a resolution of
$24$~arcsec. The Stokes $I$ FOV is shown in Fig.~\ref{SCG1-PBCOR}.

\begin{figure*}
\centering
\includegraphics[clip=true, trim=0.8cm 0.8cm 0cm 0cm, width=8.5cm]{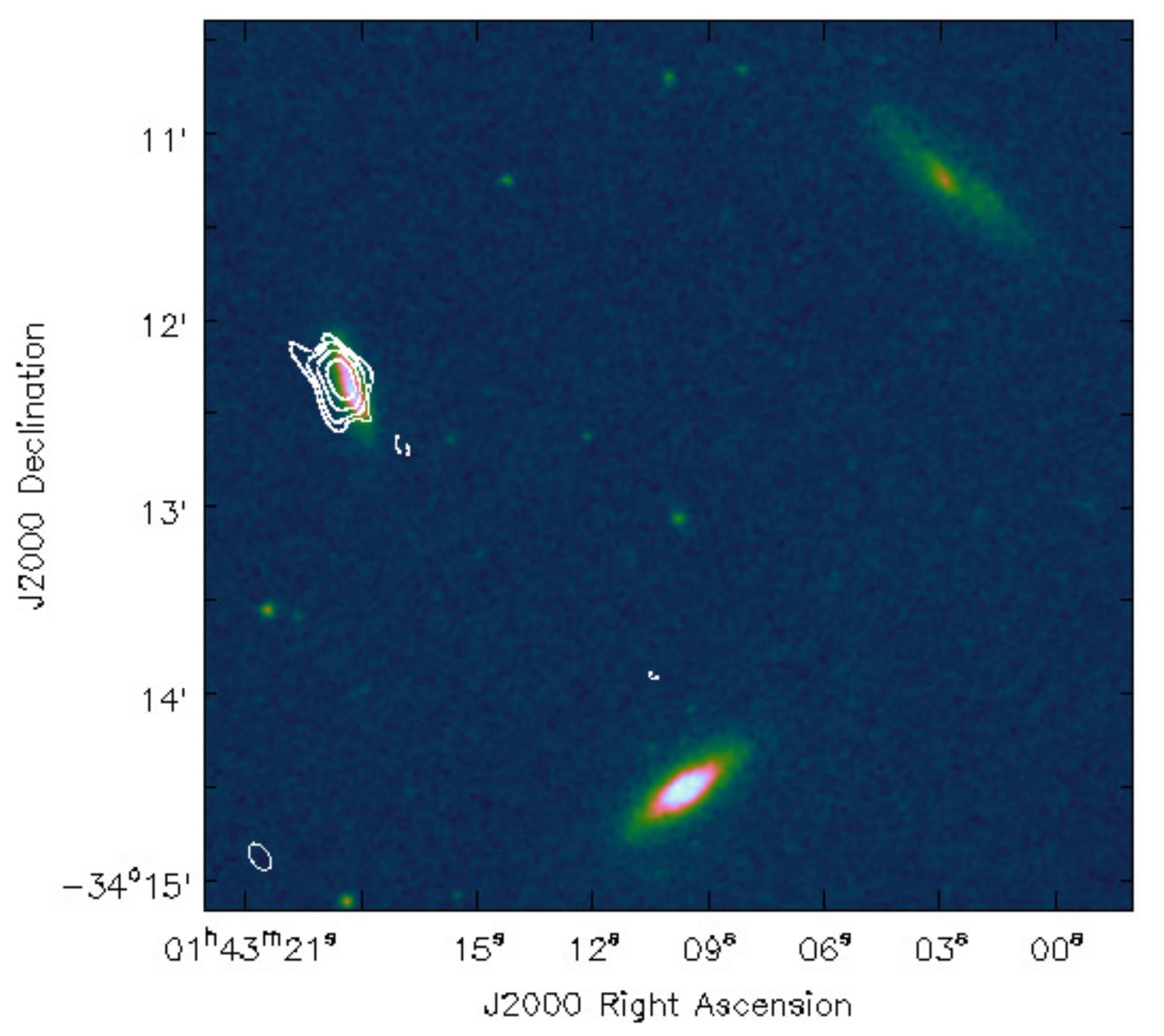}
%
%
\hspace{0pt}\includegraphics[clip=true, trim=0.9cm 0.9cm 0cm 0cm, width=8.8cm]{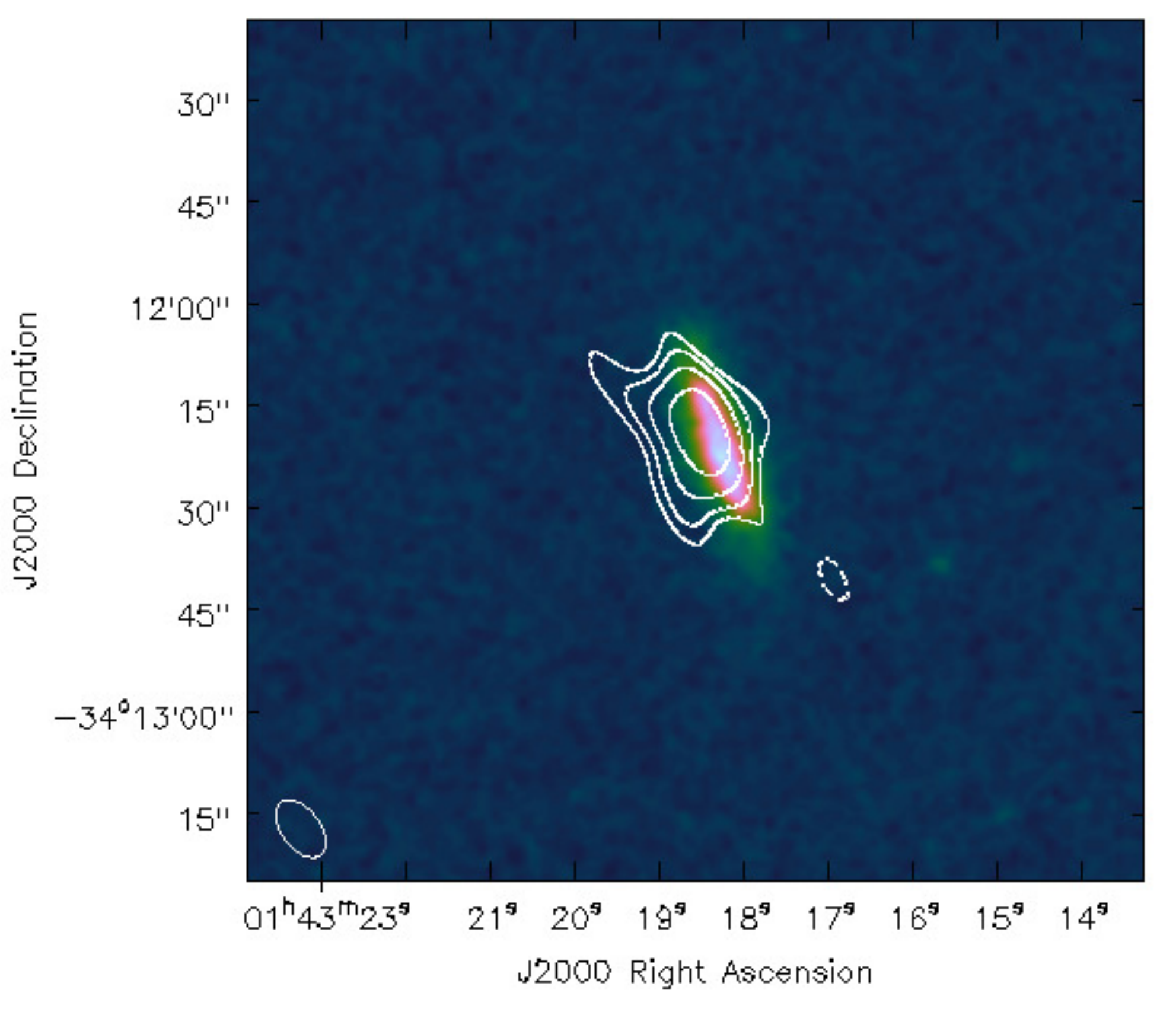}
\caption{Left: Stokes $I$ contours at 610~MHz overlaid on the near-IR
DSS-2 image of the region near ESO~0353$-$G036 in the field of
USCG~S063. The galaxies IC~1722 and IC~1724, which are not detected at
610~MHz, are visible to the northwest and south respectively. The
contours are at ($-2$, $-1$, 1, 2, 4, 8, 16) $\times$ the off-source
$4\sigma$ level, where $\sigma$ is 270~$\muup$Jy~beam$^{-1}$. The
610~MHz image is at full resolution of $9\farcs7 \times 5\farcs6$ with a
PA of $37^{\circ}$. Right: The image is the same as to the left, but zoomed
in towards ESO~0353$-$G036. The synthesised beam is shown in the bottom left.}\label{SCG1-galaxy}
\end{figure*}

At 610~MHz, galaxies IC~1724, IC~1722, IRAS~F01415--3433, and ESO~0353--G039 were all undetected in Stokes $I$ to a $5\sigma$ upper limit of $<1.35$~mJy. The galaxy ESO~0353--G036 was
detected and found to have a peak brightness of 15.4~mJy~beam$^{-1}$
(as shown in Fig.~\ref{SCG1-galaxy}). The peak of radio emission from ESO~0353--G036 appears to be offset from the emission in the near-IR, with the
offset directed towards IRAS~F01415--3433 and ESO~0353--G039 (towards the east). The offset emission is also extended by two synthesised beams towards the
east, and reaches beyond the extent of the disk that is visible at
near-IR wavelengths. Further higher resolution observations will be required to identify the cause of this emission.

None of the group galaxies were detected in polarisation. No
polarisation was detected from ESO~0353--G036 to a $3\sigma$ upper limit of
$<0.4$\%.

\begin{figure*}
\centering
\includegraphics[clip=true, trim=0.1cm 1.1cm 0cm 0cm, width=12cm]{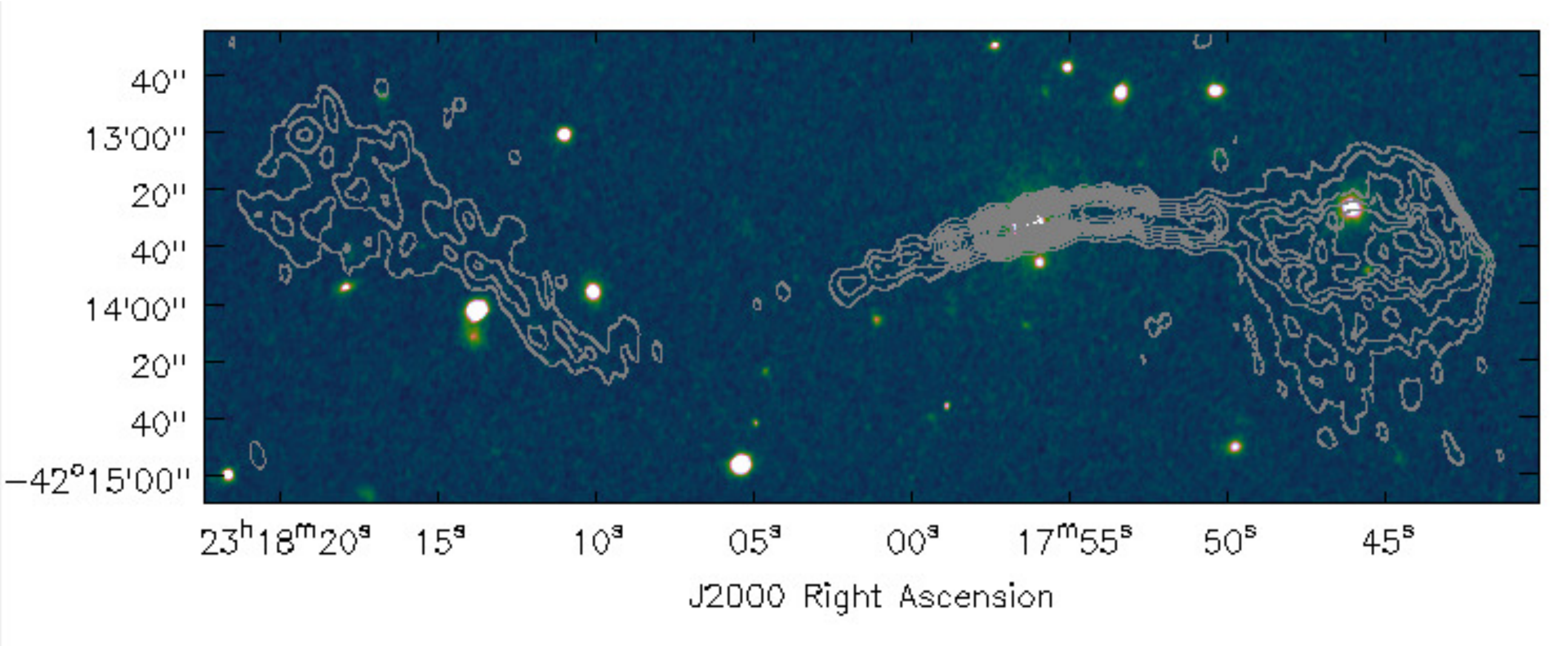}\\
\bigskip
\includegraphics[clip=true, trim=0.1cm 1.1cm 0cm 0cm, width=12cm]{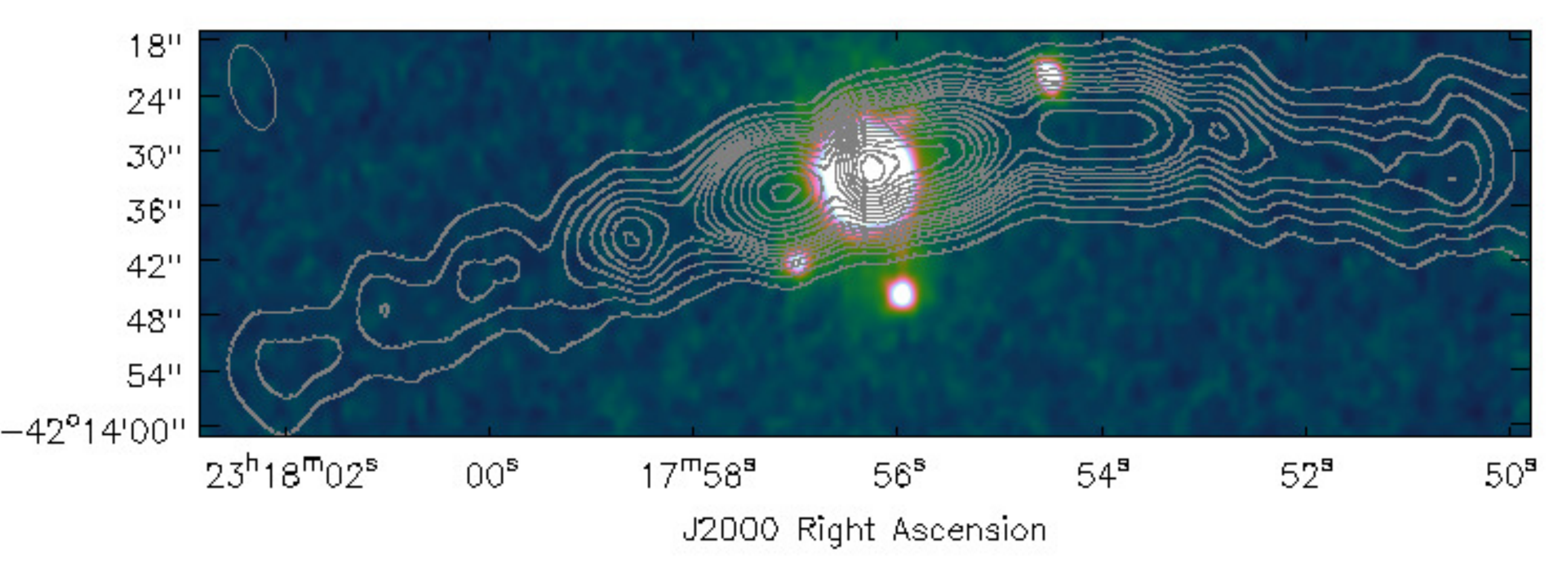}
\caption{Top: Stokes $I$ contours at 610~MHz overlaid on the near-IR DSS-2
image of the FR-I source J2317.9$-$4213 in the field of the Grus Quartet. The contours
are at ($-2$, $-1$, 1, 2, 3\ldots8, 10, 12\ldots28) $\times$ the
off-source $3\sigma$ level, where $\sigma$ is 185~$\muup$Jy~beam$^{-1}$.
The 610~MHz image is at full resolution of $9\farcs5 \times 4\farcs6$
with a PA of $16^{\circ}$. The synthesised beam is shown in the bottom left. Bottom: Similar to the top image, but zoomed in towards the centre of
J2317.9$-$4213 and with the synthesised beam shown in the top left.}\label{AGN}
\end{figure*}

\begin{figure*}
\centering
\includegraphics[clip=false, trim=0cm 0cm 0cm 0cm, width=15cm]{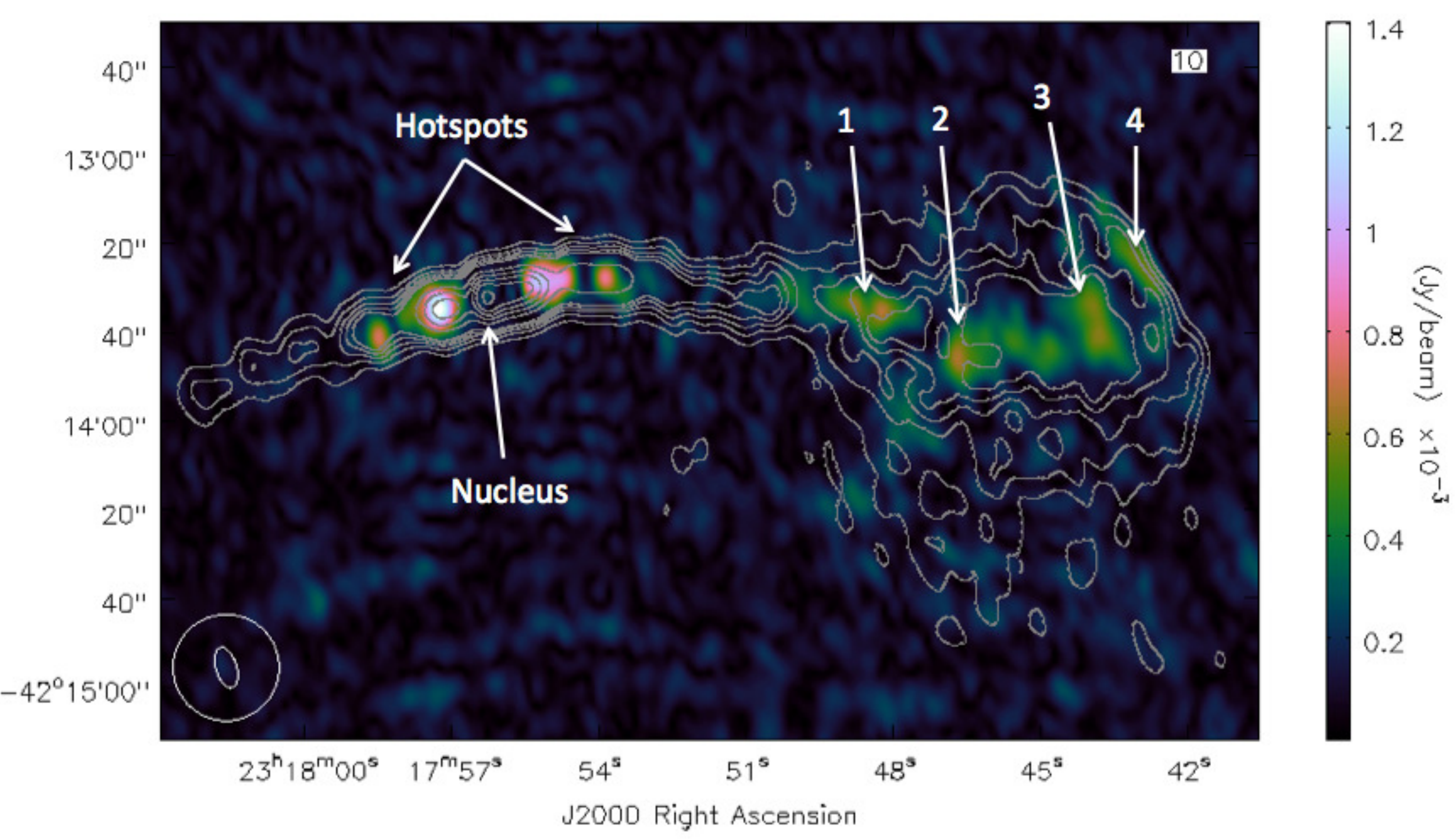}
\caption{Stokes $I$ contours at 610~MHz overlaid on the $P$ image at a Faraday depth of $10$~rad~m$^{-2}$ for the FR-I source J2317.9$-$4213 in the field of the Grus Quartet.
The grayscale is in units of mJy~beam$^{-1}$. The contours are at ($-2$,
$-1$, 1, 2, 3, 4, 5, 10, 14, 18, 22, 28) $\times$ the off-source
$3\sigma$ level, where $\sigma$ is 185~$\muup$Jy~beam$^{-1}$. The Stokes
$I$ contours are at full resolution of $9\farcs5 \times 4\farcs6$
with a PA of $16^{\circ}$. The $P$ image has a resolution of $24$~arcsec, and has not been corrected for the
effects of the primary beam or for Rician bias. The synthesised beams are shown in the bottom left.
}\label{AGNpol}
\end{figure*}

\section{Other Extended Sources}\label{S:nongroup}

Three spatially extended sources were detected in the field of the Grus
quartet that are unrelated to the group galaxies themselves. These three
radio sources J2317.9$-$4213, MCG$-$07$-$47$-$031, and PKS~2316$-$429
have all been previously detected but are largely unstudied
\citep{1980BAAS...12..804E, 1981MNRAS.194..693L, 1998AJ....115.1253P,
2009A&A...495..691M}. In this section, these individual sources are
described in further detail. As we shall show, J2317.9$-$4213 resembles
an FR-I radio source, MCG$-$07$-$47$-$031 is classified as a blazar, and
PKS~2316$-$429 is likely an FR-II X-shaped radio source. This allows a
glimpse into the polarisation properties of these different types of
sources at 610~MHz.

\subsection{J2317.9$-$4213}

The complex radio galaxy J2317.9$-$4213 is located $\sim 1$~arcmin to the west of NGC~7590 at $23^{\rm h} 17^{\rm m} 55\fs7$, $-42^{\circ} 13'
29''$. It has possibly been detected as an X-ray source, indicating the
presence of an AGN \citep{1980BAAS...12..804E}. It is well-resolved,
with the two jets/lobes projecting along an axis that is approximately aligned
east--west, as shown in Fig.~\ref{AGN} and \ref{AGNpol}. The spectrum, $\alpha_{610}^{1425}$, was found to be
$0.43\pm0.15$ at the location of the AGN, and $0.84\pm0.26$ and
$0.56\pm0.12$ in the brightest emission from the east and west jet
respectively. The jet to the east curves away to the north relative to
the approximate east--west axis as seen in projection. Such distortions
are commonplace in FR-I sources \citep[e.g.][]{1992ersf.meet..307L,
2002MNRAS.330..609D, 2007AJ....133.2097C}. This suggests interaction
between the surrounding environment to the east and the expanding jet --
resulting from either ram pressure or a denser ambient medium
\citep[e.g.][]{1984RvMP...56..255B, 1986ApJ...307...73B,
1998MNRAS.299..357B}.

In the near-IR DSS images is an apparent counterpart source coincident
with the brightest 610~MHz Stokes $I$ emission. This counterpart is shown in Fig.~\ref{AGN}. The counterpart has $z=0.056$ and is therefore in the background of the galaxy group, with
the emission from the lobes extending across 43~kpc. The general
morphology, in combination with the presence of low brightness regions
towards the periphery is indicative of the FR-I class. The 610~MHz
polarisation is brighter in the western jet and the contrasting lack of
polarised emission to the east is likely a consequence of the
Laing--Garrington effect \citep[e.g.][]{1988Natur.331..147G}. This
suggests the eastern jet is located more distantly along the
line-of-sight and therefore subject to stronger Faraday depolarisation.

No polarisation was detected from the AGN, to a $3\sigma$ upper limit of
$<0.2$\%. However, there are two pairs of polarised hotspots located on
both the east and west side of the nucleus, as shown in
Fig.~\ref{AGNpol}. These hotspots are approximately equidistant from
each other and the associated AGN. The hotspots to the east have
fractional polarisations of $5.2\pm0.4$\% and $4.0\pm0.3$\%, and the
hotspots to the west are polarised to $3.2\pm0.3$\% and $2.2\pm0.2$\%.
The lesser of each of these values is associated with the hotspot that
is in closer proximity to the AGN. These values also show that the
polarisation is $\sim 2$\% higher on the east side of the nucleus.

Moving from the hotspot furthest to the east, through to the west, the
FDs extracted from the datacubes are $15.9\pm1.3$~rad~m$^{-2}$,
$12.7\pm0.5$~rad~m$^{-2}$, $22.1\pm0.6$~rad~m$^{-2}$, and
$27.4\pm1.1$~rad~m$^{-2}$. The FDs are clearly higher on the western
side. Relative to the outer polarised hotspots, the pair nearest to the
AGN have lower FDs. The polarised hotspots are associated with regions
that also show an increase in total intensity, and may originate from
strong shocks along the boundary of the jet outflow. A strong shock
would compress and order the magnetic field, leading to an increase in
the linear polarisation fraction. If this is the case, the polarisation
fractions would mildly suggest a denser ambient medium to the east in
near proximity to the AGN. The combination of the observed polarisation
fractions and Faraday depths of these hotspots support two scenarios
near to the AGN, that are not mutually exclusive: either the magnetic
field strength is weaker, or alternatively there is a combination of
increased turbulence, that increases the random component of the
magnetic field, together with lower electron density.

The western jet can be approximately split into four regions of
polarised emission, as indicated in Fig.~\ref{AGNpol}. The FD of these four regions of extended
polarisation in the western jet are (moving from east to west):
$-11.7\pm2.4$~rad~m$^{-2}$, $-2.0\pm1.1$~rad~m$^{-2}$,
$-10.9\pm1.1$~rad~m$^{-2}$, and $+4.7\pm1.4$~rad~m$^{-2}$. There appears
to be a large scale change of heading of the magnetic field along the
line of sight (as shown by the predominantly negative FD) relative to that near to the nucleus (which has positive FD). The polarised region (number 4) furthest to the northwest, and at the outer periphery of jet, again has a positive FD. This suggests another change in the heading of the magnetic field along the line of sight relative to the rest of the western jet. The morphology of this region is
indicative of shocked material causing an increase in the ordering of
the magnetic field, with the polarised emission laying along the western
edge.

\subsection{MCG$-$07$-$47$-$031}

\begin{figure}
%
\includegraphics[clip=true, trim=0.2cm 0.9cm 0cm 0cm,
    height=6.0cm]{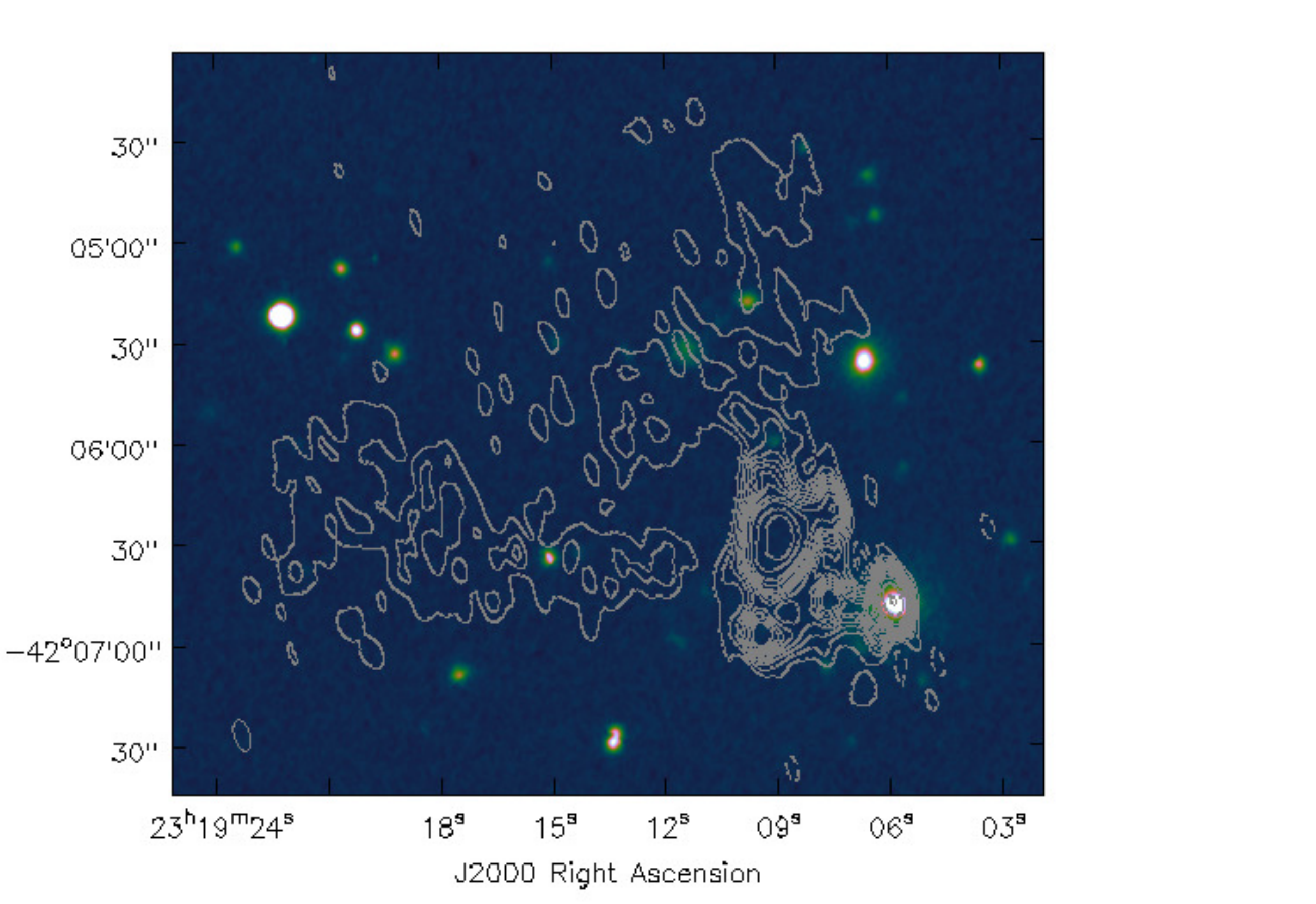}\\
\bigskip
%
\includegraphics[clip=true, trim=0.2cm 0.7cm 0cm 0cm,
    height=5.7cm]{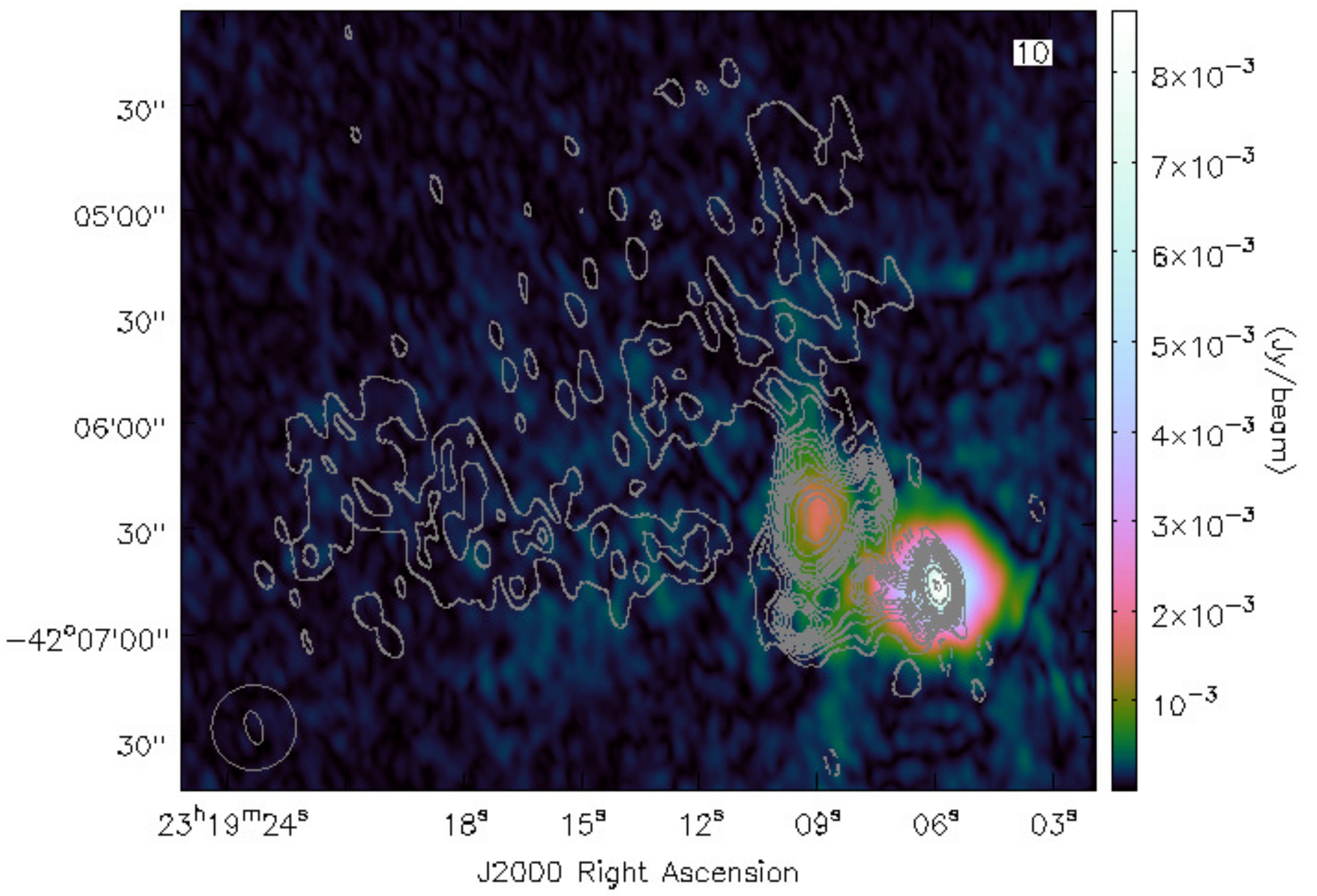}
\caption{Top: Stokes $I$ contours at 610~MHz overlaid on the near-IR
DSS-2 image of the BL Lac MCG$-$07$-$47$-$031 in the field of the Grus Quartet. The
contours are at ($-2$, $-1$, 1, 2, 3\ldots8, 10, 12, 16, 20, 32, 64,
128) $\times$ the off-source $3\sigma$ level, where $\sigma$ is
555~$\muup$Jy~beam$^{-1}$. The 610~MHz image is at full resolution of
$9\farcs5 \times 4\farcs6$ with a PA of $16^{\circ}$. Bottom: Stokes $I$
contours at 610~MHz overlaid on the $P$ image of
MCG$-$07$-$47$-$031 at a Faraday depth of $10$~rad~m$^{-2}$. The $P$ image has a resolution of $24$~arcsec, and
has not been corrected for the effects of the primary beam or for Rician
bias. The synthesised beams are shown in the bottom left.}\label{BLLAC}
\end{figure}

The blazar or BL Lac, MCG$-$07$-$47$-$031, is located $\sim 10$~arcmin
to the north of galaxy NGC~7599 at $23^{\rm h} 19^{\rm m} 5\fs8$,
$-42^{\circ} 06' 49''$. This bright AGN appears to have a near-IR
counterpart in the DSS images, which is located at $z=0.055$, or
$228$~Mpc away. The identified counterpart in J2317.9$-$4213 is located
at $z=0.056$. The similar redshifts of these two sources may indicate
that they are part of the same background cluster. The Stokes $I$, near-IR data, and polarised intensity at 610~MHz are shown in Fig.~\ref{BLLAC}.

Near the core in Stokes $I$, there are at least three other point-like
regions of emission that appear associated with the blazar -- although
foreground/background sources cannot be ruled out. At the faintest
levels in the images, there is a $3\sigma$ detection of two jet-like
features leading from the AGN -- one directly north, and another to the
east. There is also some indication of a bridge of emission connecting
the two jets.

The nucleus was found to have a flat spectrum typical of a blazar of
$\alpha_{610}^{843}=0.29\pm0.25$, and is detected in linear polarisation
at this epoch with a fraction of $7.03\pm0.11$\%. A point-like source to
the east of the nucleus is also polarised to $1.88\pm0.05$\%, although instrumental effects are anticipated to be on the order of at most $\sim
2.5$\% at $23\farcm7$ from the phase-centre. Nevertheless, this polarised emission is likely real -- with the peak in polarised intensity having a non-zero FD$=2.5\pm0.6$~rad~m$^{-2}$. The nucleus itself has a FD
of $6.5\pm0.1$~rad~m$^{-2}$. No polarisation was detected from the jets
to a $3\sigma$ upper limit of $<0.7$\%.

\subsection{PKS~2316$-$429}

\begin{figure}
%
\includegraphics[clip=true, trim=0.9cm 1.1cm 0cm 0cm,
    height=6.5cm]{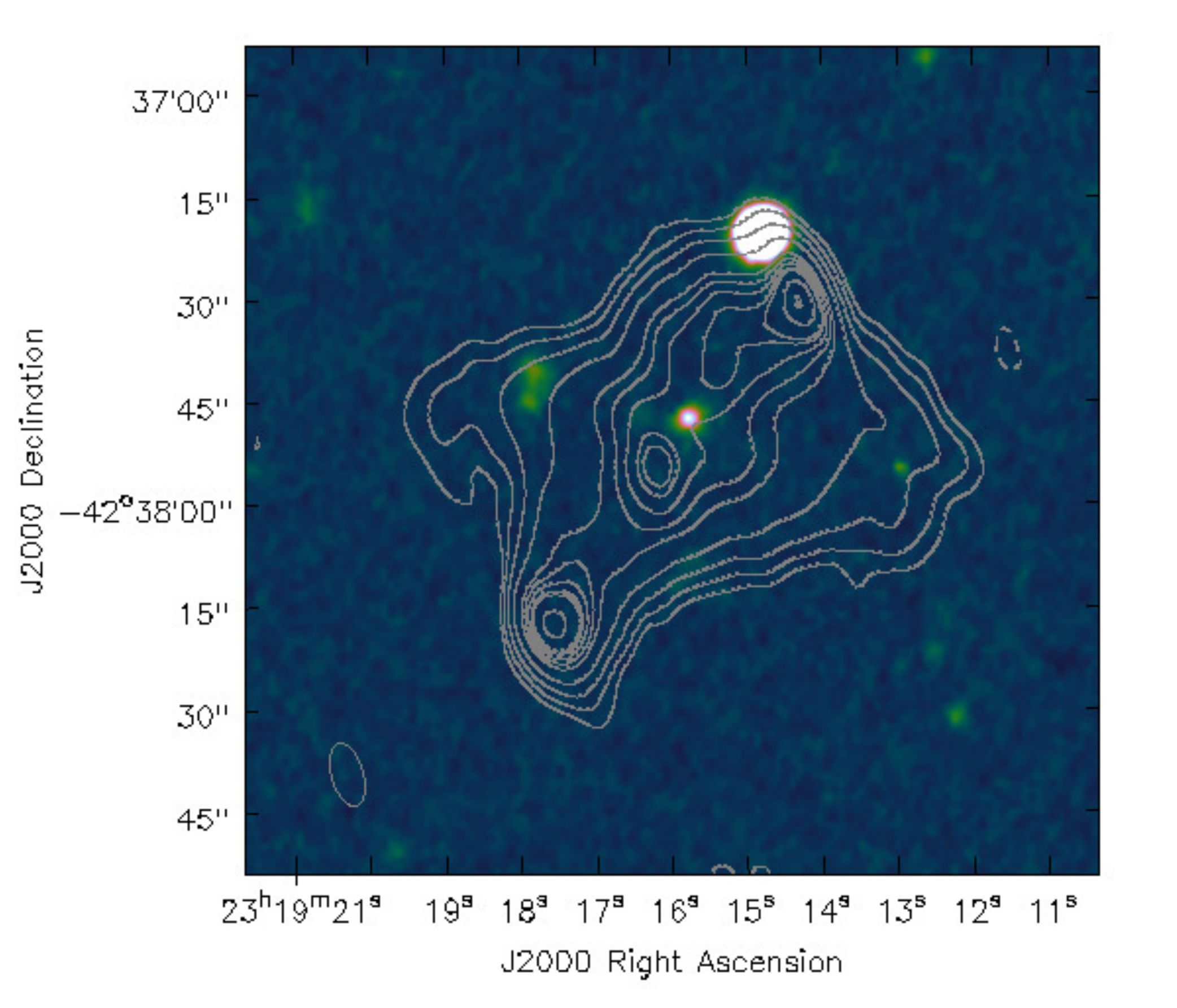}\\
\bigskip
%
\includegraphics[clip=true, trim=0.7cm 0.7cm 0cm 0cm,
    height=6.25cm]{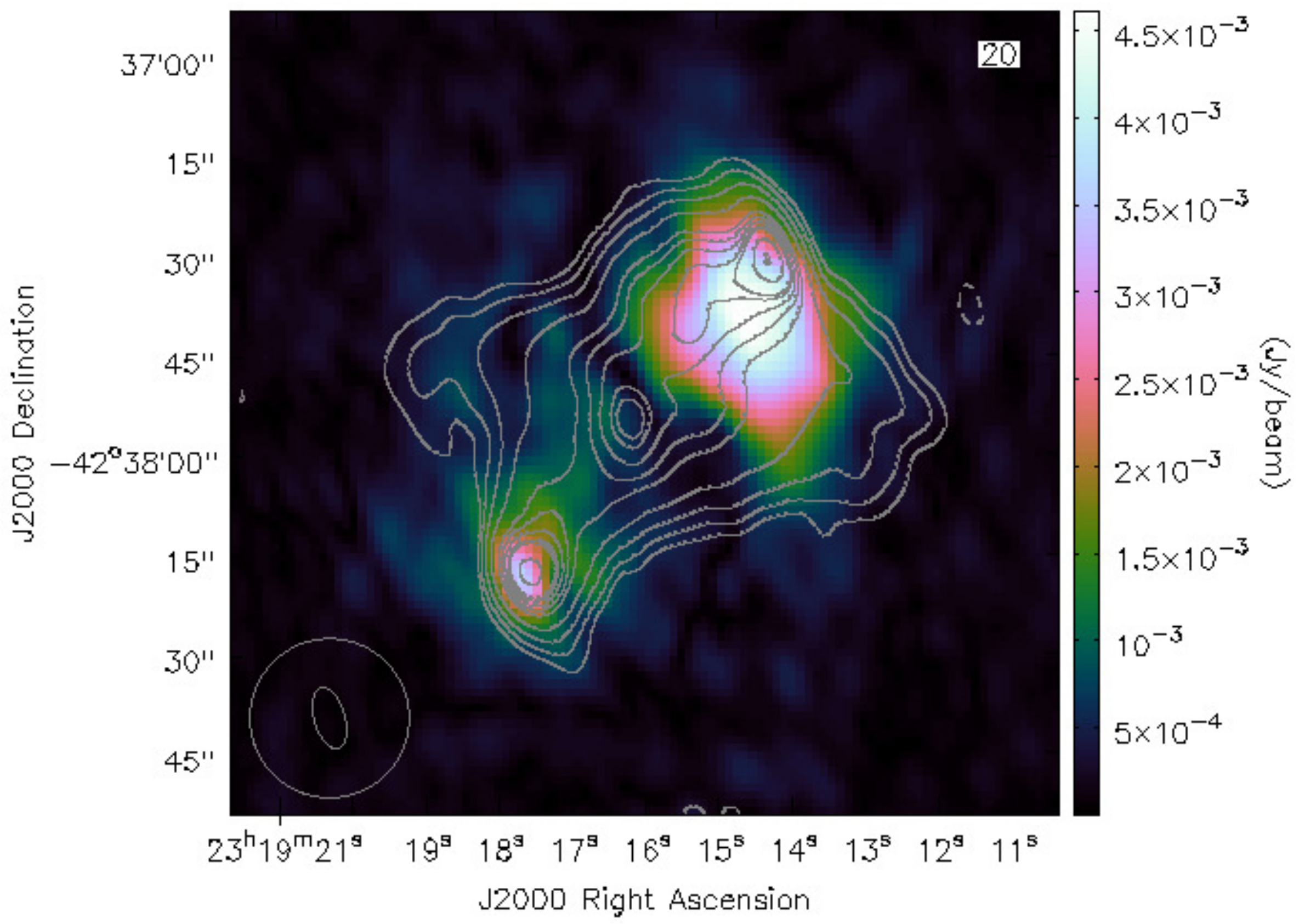}
\caption{Top: Stokes $I$ contours at 610~MHz overlaid on the near-IR
DSS-2 image of the X-shaped radio source PKS~2316$-$429 in the field of the Grus Quartet. The
contours are at ($-2$, $-1$, 1, 2, 4, 8, 16, 20, 24, 28, 30, 40, 50)
$\times$ the off-source $4\sigma$ level, where $\sigma$ is
225~$\muup$Jy~beam$^{-1}$. The 610~MHz image is at full resolution of
$9\farcs5 \times 4\farcs6$ with a PA of $16^{\circ}$. Bottom: Stokes $I$
contours at 610~MHz overlaid on the $P$ image of
PKS~2316$-$429 at a Faraday depth of $20$~rad~m$^{-2}$. The $P$ image has a resolution of $24$~arcsec, and has
not been corrected for the effects of the primary beam or for Rician
bias. The synthesised beams are shown in the bottom left.}\label{SE-SCG2}
\end{figure}

This previously unclassified radio source is located $\sim 25$~arcmin
south of galaxy NGC~7599 at $23^{\rm h} 19^{\rm m} 15\fs9$, $-42^\circ 37'
53''$. The radio morphology has a central unresolved core, with two
bright hotspots to the northwest and southeast periphery. Diffuse
emission extending along the east--west is also detected, giving the
source an X-shaped morphology. The Stokes $I$, near-IR data, and polarised intensity at 610~MHz are shown in Fig.~\ref{SE-SCG2}.

Two near-IR sources enclosed within the Stokes $I$ emission are feasibly
in the background and unrelated to the radio source, as their location
bears no strong relevance to the radio structure. Nevertheless, one of
these sources is near to the approximate intersection of the X-shaped
structure, and could be an associated AGN -- although no redshift data
was available. Note that the identified counterpart in J2317.9$-$4213 is
located at $z=0.056$, and the counterpart in MCG$-$07$-$47$-$031 is
located at $z=0.055$. The similar redshifts of these two sources, when
also considered together with PKS~2316$-$429 and the similar angular
size of all three radio sources, may indicate that they are all part of
the same background cluster. Nevertheless, this is not clear without
redshift information for PKS~2316$-$429.

The PKS~2316$-$429 radio source is also detected in the SUMSS, and
comparison with the 610~MHz data shows the source to have a non-thermal
spectrum with $\alpha_{610}^{843}=0.57\pm0.13$. There are two peaks in
linear polarisation that are offset from the two total intensity
hotspots. The polarisation fraction determined from the FDF at the
position of the peak polarised emission is $3.86\pm0.07$\% and
$6.7\pm0.3$\% to the northwest and southeast respectively. Direction-dependent effects are estimated to be at most $\sim 2.0$\%.
The FD is similar for both polarisation peaks, being
$15.39\pm0.15$~rad~m$^{-2}$ and $16.56\pm0.19$~rad~m$^{-2}$ to the
northwest and southeast. No polarisation was detected from the core to a
$3\sigma$ upper limit of $<0.6$\%.

The non-thermal spectrum, polarisation properties, and radio morphology
-- particularly the bright outer hotspots -- is typical of an FR-II
type, X-shaped radio galaxy with both a set of active lobes and wings.
The GMRT has previously been used to study such objects
\citep{2007MNRAS.374.1085L,2008arXiv0810.0941L}. Only $\sim$100 candidate sources at most
have been identified to date, and even this is likely an overestimated
sample that includes Z- and S-shaped distortions that are common in FR-I
sources \citep[e.g.][]{1992ersf.meet..307L, 2002MNRAS.330..609D,
2007AJ....133.2097C}. A frequent explanation for the unusual morphology
is that the central supermassive black hole underwent a recent
realignment \citep[e.g.][]{2002Sci...297.1310M}.

The complementary SUMSS data does not have sufficient resolution to
measure any spatial spectral index variations in this X-shaped galaxy.
Future observations could allow for the spectral index of the active and
the winged lobes to be measured -- the active lobes typically have
$\alpha \approx 0.7$ and the wings $\alpha \approx 1.2$, and the aged
electron population can reveal which of the jets is from the more recent
period of AGN activity \citep{2002MNRAS.330..609D}. Nevertheless, the
bright hotspots and coincident polarisation are typical of termination
shocks as the jet fluid encounters approximately stationary lobe
material -- indicating that these are the active lobes.

\section{Discussion and Conclusions}\label{S:discussion}

We have shown that it is possible to calibrate the GMRT's
full-polarisation mode for the use of spectropolarimetry, in
order to investigate some aspects of cosmic magnetism. In addition, our characterisation of the wide-field polarisation beam allows us to be confident of polarisation measurements within an appropriate portion of the FOV. The application
of RM Synthesis to GMRT data requires consideration of a large number of
systematics, including the instrumental polarisation, instrumental
time-stability, quality of $uv$-data calibration, ionospheric Faraday
rotation, polarisation angle corrections across the observing bandwidth,
and the wide-field response. However, we have shown that all of these
calibration issues can be addressed, resulting in a sensitivity of a few tens of $\muup$Jy in Faraday-space at 610~MHz.

Nevertheless, there are a number of limitations for GMRT polarisation
data that restrict the instrument's current use for high-dynamic range
polarimetric observations. The on-axis instrumental polarisation is both
large and highly frequency-dependent at 610~MHz. The rapid wrapping of
phase with frequency limits the quality of polarisation calibration.
Furthermore, the fact that some antennas have high leakage requires the
use of a non-linear model in order to separate the source and
instrumental polarisation. Nevertheless, a full non-linearised model is
not currently available in commonly used data reduction packages, \citep[see][for further detail]{mythesis}. It has been possible to make progress
with GMRT polarimetry by removing antennas with instrumental
polarisation $\ge$15\%, thereby eliminating antennas for which a
small-leakage approximation breaks down. The loss of antennas that are
located in the outer arms of the array frequently leaves significant
gaps in the $uv$-coverage. This requires tapering of the data in the
$uv$-plane, which reduces the highest resolution that is available.
Nevertheless, the major upgrade that is currently underway will
essentially revolutionise GMRT polarimetry, as the better quality
receivers that are currently being installed will likely have completely
different polarisation properties.

The GMRT has been used for both deep surveys at 610~MHz
\citep[e.g.][]{2007MNRAS.376.1251G, 2008MNRAS.383...75G,
2008MNRAS.387.1037G, 2009MNRAS.397.1101G, 2010BASI...38..103G}  and
large-area surveys such as the TIFR GMRT Sky Survey (TGSS) at
150~MHz\footnote{\texttt{http://tgss.ncra.tifr.res.in/}}. The TGSS is
planned to detect up to two million sources in total intensity at
150~MHz. The resolution and sensitivity of the GMRT at low radio
frequencies means that polarimetric radio surveys have the potential to
be complementary to planned surveys with facilities such as ASKAP that
explore a similar parameter space at higher frequencies and more
Southern declinations \citep[e.g.][]{2008ExA....22..151J}.

Future observations with the polarisation mode of the GMRT could make
useful predictions of the number density of polarised sources observable
with the SKA and its pathfinders, and opens up the possibility to model
magnetic fields and their evolution in populations of faint radio
sources \citep[e.g.][]{2007mru..confE..69S, 2009ASPC..407...12T}.
Furthermore, it has been argued by \citet{2012A&A...543A.113B} that a
combination of GMRT, LOFAR, and JVLA data will allow for magnetic
structures at intermediate scales to be recognised -- comparable to the
range of recognisable scales in Faraday space for SKA data. This would
allow for much stronger constraints to be placed on the prevailing
depolarisation mechanisms at low frequencies, allow for improved
estimates of polarised source counts at low frequencies, and demonstrate
the feasibility of using next-generation facilities, such as the SKA,
for observing the RM-grids that will be essential for understanding
cosmic magnetism \citep{2004NewAR..48.1003G}.

\section*{Acknowledgments}

We thank the staff of the GMRT that made these observations possible.
The GMRT is run by the National Centre for Radio Astrophysics of the
Tata Institute of Fundamental Research. We are grateful to Julia Riley for helpful comments on early versions of the manuscript. We thank the anonymous referee for useful comments on the paper. J.S.F.\ has been supported by the Science and Technology Facilities Council, and acknowledges the support of the Australian Research Council through grant DP0986386.


\bsp

\label{lastpage}

\end{document}